\documentclass[useAMS,usenatbib]{mn2e}

\usepackage{graphicx}
\usepackage{graphics}
\usepackage{natbib}
\usepackage{amsmath}
\usepackage{subfig}

\title[HST optical polarimetry of the Vela pulsar \& nebula]
  {HST optical polarimetry of the Vela pulsar \& nebula}
\author[P. Moran et al.]
  {P.~Moran,$^1$\thanks{E-mail: p.moran4@nuigalway.ie}  
  R. P.~Mignani,$^{2,3}$ A.~Shearer$^1$\thanks{E-mail: andy.shearer@nuigalway.ie}    %
\\
  $^1$Centre for Astronomy, School of Physics, National University of Ireland Galway, University Road, Galway, Ireland\\
  $^2$INAF - Istituto di Astrofisica Spaziale e Fisica Cosmica Milano, via E. Bassini 15, 20133, Milano, Italy\\
  $^3$Kepler Institute of Astronomy, University of Zielona G\'ora, Lubuska 2, 65-265, Zielona G\'ora, Poland\\
  }
  
\date{Released 2014 Xxxxx XX}

\pagerange{\pageref{firstpage}--\pageref{lastpage}} \pubyear{2014}

\def\LaTeX{L\kern-.36em\raise.3ex\hbox{a}\kern-.15em
    T\kern-.1667em\lower.7ex\hbox{E}\kern-.125emX}

\begin{document}

\label{firstpage}

\maketitle


\begin{abstract}
Polarisation measurements of pulsars offer an unique insight into the geometry of the emission regions in the neutron star magnetosphere. 
Therefore, they provide observational constraints on the different models proposed for the pulsar emission mechanisms. Optical polarisation data of the Vela pulsar was obtained from the {\em Hubble Space Telescope} ({\em HST}) archive. The data, obtained in two filters (F606W; central wavelength = 590.70 nm, and F550M; central wavelength = 558.15 nm), consists of a series of observations of the pulsar taken with the {\em HST}/Advanced Camera for Surveys (ACS) and covers a time span of 5 days. This data have been used to carry out the first high-spatial resolution and multi-epoch study of the polarisation of the pulsar. We produced polarisation vector maps of the region surrounding the pulsar and measured the degree of linear polarisation (P.D.) and the position angle (P.A.) of the pulsar's integrated pulse beam. 
We obtained $\rm P.D.=8.1\%\pm0.7\%$ and $\rm P.A.=146.3\degr\pm2.4\degr$, averaged over the time span covered by these observations. These results 
not only confirm those originally obtained by \citeauthor{Wagner00} and \citeauthor{Mignani07}, both using the Very Large Telescope (VLT), but are of  greater precision. Furthermore, we confirm that the P.A. of the pulsar polarisation vector is aligned with the direction of the pulsar proper-motion. The pulsar wind nebula (PWN) is undetected in polarised light as is the case in unpolarised light, down to a flux limit of 26.8 magnitudes arcsec$^{-2}$.
\end{abstract}




\section{Introduction}

Strong polarisation is expected when the pulsar optical emission is generated by synchrotron radiation. \citet{Shklovsky53} suggested that the continuous optical radiation from the Crab Nebula was due to synchrotron radiation. This was later confirmed by \citet{Dombrovsky54} and \citet{Vashakidze54} who found that the optical radiation was polarised. Incoherent synchrotron emission follows a simple relationship between its polarisation profile and underlying geometry. Hence, optical polarisation measurements of pulsars provide an unique insight into the geometry of their emission regions, and therefore observational constraints on the theoretical models of the emission mechanisms. From an understanding of the emission geometry, one can limit the competing models for pulsar emission, and hence understand how pulsars work - a problem which has eluded astronomers for almost 50 years.

Polarimeters are sensitive in the optical, but the majority of pulsars are very faint at these wavelengths with V $\ge$ 25 \citep{Shearer08}. Polarimetry in the very high-energy domain, X-ray and $\gamma$-ray, using instruments on board space telescopes, is of limited sensitivity. So far, detailed results have only been reported for the Crab pulsar \citep{Weisskopf78,Dean08,Forot08}. Although the number of pulsars detected in the optical is growing \citep{Mignani11}, only five pulsars have had their optical polarisation measured; Crab \citep{Wampler69, Kristian70, Smith88, Slowikowska09, Moran13}, Vela \citep{Wagner00, Mignani07}, PSR\, B0540--69 \citep{Middleditch87,Chanan90,Wagner00,Mignani10} PSR\, B0656+14 \citep{Kern03}, and PSR\, B1509--58 \citep{Wagner00}. Nonetheless, the optical currently remains invaluable for polarimetry in the energy domain above radio photon energies. The Crab pulsar, being the brightest optical pulsar with V $\approx 16.8$ \citep{Nasuti96}, has had several measurements of its optical polarisation, 
both phase-averaged and phase-resolved.
Hence, polarisation studies of more pulsars are needed to better understand and constrain the polarisation properties of pulsars and search for possible correlations with their intrinsic parameters (e.g. age, magnetic field strength).

The Vela pulsar is a young energetic pulsar, with characteristic age $\tau$ $\sim$ 11 kyr and spin-down power $\dot{E}$ $\sim$ 6.9 $\times$ 10$^{36}$ erg s$^{-1}$,  associated with the Vela supernova remnant \citep{Large68}. It is the third brightest optical pulsar (V $\sim$ 23.6) \citep{Mignani01} and is located at a distance of $\approx$ 290 pc \citep{Caraveo01}. It powers the nearest bright X-ray pulsar wind nebula (PWN), a plerion of positrons and electrons that is created by the confinement of the pulsar wind by its surrounding supernova remnant. For detailed reviews of PWNe see \citet{Gaensler06}, \citet{Kargaltsev08}, and \citet{Kargaltsev13}. The Vela PWN consists of a double arc structure \citep{Helfand01, Pavlov01b, Pavlov03} and an X-ray jet \citep{Markwardt95}, aligned along the axis of symmetry of the arcs. The bilateral structure of the PWN is remarkably similar to that seen by {\em Chandra} observations of the Crab nebula \citep{Weisskopf00}. Radio observations, using the Australian Telescope Compact Array (ATCA), have revealed a highly polarised (60\%) and extended radio nebula with symmetric morphology surrounding the pulsar and X-ray nebula \citep{Dodson03a}. However, to date no detection of the optical counterpart of the Vela PWN has been reported, despite deep observations with the {\em Hubble Space Telescope} ({\em HST}) \citep{Mignani03}.

Here, we present the results of recent polarisation observations of the Vela pulsar field carried out with the Advanced Camera for Surveys (ACS) of the {\em HST}.
The purpose of this work is four fold. Firstly, we wanted to independently confirm the polarisation measurements for the Vela pulsar originally reported by \citet{Wagner00} and \citet{Mignani07}, both obtained with the Very Large Telescope (VLT), and improve their accuracy, obtaining the highest-significance polarisation measurement of the pulsar. 

Secondly, we wanted to check for possible short-term variability of the pulsar polarisation, an experiment that can be ideally conducted with the {\em HST} being unaffected by the variable polarisation background of the Earth's atmosphere. Thirdly, since we expect that fine structures within the X-ray PWN are strongly polarised in the optical, like for the Crab, we examined the {\em HST}/ACS images to search, for the first time,  for polarised optical emission from the Vela PWN. Detection of the PWN in the optical, together with the radio and X-ray data, will help establish the properties of the relativistic pulsar wind, including its energetics, magnetic field structure, spatial evolution and interaction with the ambient medium. As shown, e.g. by the cases of the Crab pulsar \citep{Moran13} and PSR\, B0540--69 \citep{Mignani10}, it is difficult to determine the polarisation for pulsars embedded in highly-polarised environments, such as those of the surrounding PWNe and supernova remnants (SNRs). So, in order to determine the Vela pulsar's polarisation profile, we need to know the level of background polarisation induced by the surrounding PWN/SNR environment. Therefore, the fourth purpose of this work is to accurately map the polarisation of the innermost part of the Vela SNR, around the pulsar position.

Our work  will then act as a guideline for future phase-resolved polarisation measurements of the Vela pulsar using, e.g. the Galway Astronomical Stokes Polarimeter (GASP). This is an ultra-high-speed, full Stokes, astronomical imaging polarimeter based on the Division of Amplitude Polarimeter (DOAP)
and has been designed to resolve extremely rapid variations in objects such as optical pulsars and magnetic cataclysmic variables \citep{Kyne10,Collins13}. 


\section{Observations and Data  Analysis}

\begin{table*}
 \caption{Summary of the HST/ACS observations of the Vela SNR taken with the WFC. The filters used were F606W ($\lambda=590.70$ nm, $\Delta\lambda=250.00$ nm) and F550M ($\lambda=558.15$ nm, $\Delta\lambda=54.70$ nm).}
 \begin{tabular}{|c|c|c|l|c|c|}
  \hline
  Date & Exposure (s) & Filter & Polariser & Roll-Angle (PA\_V3) (\degr) & Pulsar Position on Chip (x,y) \\
  \hline
  2011 Feb 18  &2$\times$1362.5  &F550M  &CLEAR2L  &205.9 & 1167.83   1169.08\\
  &2$\times$1299.5  &F606W  &POL0V  & &\\
  &2$\times$1386.5  &  &POL60V & &\\
  &2$\times$1386.5  &  &POL120V & &\\
  \hline
  2011 Feb 19 &2$\times$1299.5  &F606W  &POL0V &203.9 & 1162.19   1173.87\\
  &2$\times$1386.5  &  &POL60V & &\\
  &2$\times$1386.5  &  &POL120V & &\\
  \hline
  2011 Feb 20 &2$\times$1299.5  &F606W  &POL0V &203.9 & 1162.10   1174.46\\
  &2$\times$1386.5  &  &POL60V & &\\
  &2$\times$1386.5  &  &POL120V & &\\
  \hline
  2011 Feb 21 &2$\times$1299.5  &F606W  &POL0V &204.9 & 1165.10   1171.40\\
  &2$\times$1386.5  &  &POL60V & &\\
  &2$\times$1386.5  &  &POL120V & &\\
  \hline
  2011 Feb 23 &2$\times$1299.5  &F606W  &POL0V &204.9 & 1165.29   1171.90\\
  &2$\times$1386.5  &  &POL60V & &\\
  &2$\times$1386.5  &  &POL120V & &\\
  \hline
 \end{tabular}
\end{table*}



\subsection{Observations and data reduction}

We downloaded the raw {\em HST}/ACS polarisation science frames of the Vela pulsar field from the Mikulski Archive for Space Telescopes\footnote{https://archive.stsci.edu/} (MAST). The dataset is comprised of 5 observations carried out in 5 different visits between 2011 February 18 and February 23 (Proposal ID: 12240). Observations were obtained using the Wide Field Channel (WFC) detector of the ACS. The WFC employs a mosaic of two $4096\times2048$ Scientific Imaging Technologies (SITe) CCDs, with a pixel-scale of $\sim0\farcs05$, covering a nominal field--of--view (FOV) of $\sim202\arcsec\times202\arcsec$ \citep{Pavlovsky04}. For these observations, with the polarisers in place, the FOV was $\approx102\arcsec\times102\arcsec$. The observations were obtained with three different polariser elements, POL0V, POL60V, POL120V, corresponding to  rotation angles of 0\degr, 60\degr, and 120\degr, respectively. For each epoch, the integrations were split in pairs of exposures of  1299.5 s each for the POL0V and 1386.5 s each for the POL60V and POL120V.  The filter used was the F606W ($\lambda=590.70$ nm, $\Delta\lambda=250.00$ nm) for the integrations with the polariser.  Two 1362.5 s exposures with the F550M filter  ($\lambda=558.15$ nm, $\Delta\lambda=54.70$ nm) and no polariser in place were taken on 2011 February 18. During each observation, {\em HST} pointed at the pulsar at about the same spacecraft roll angle ($\sim$ 205\degr), with the pulsar centred at roughly the same CCD position. See Table 1 for a summary of the observations.  

\begin{figure*}
\centering
\includegraphics[width=125mm]{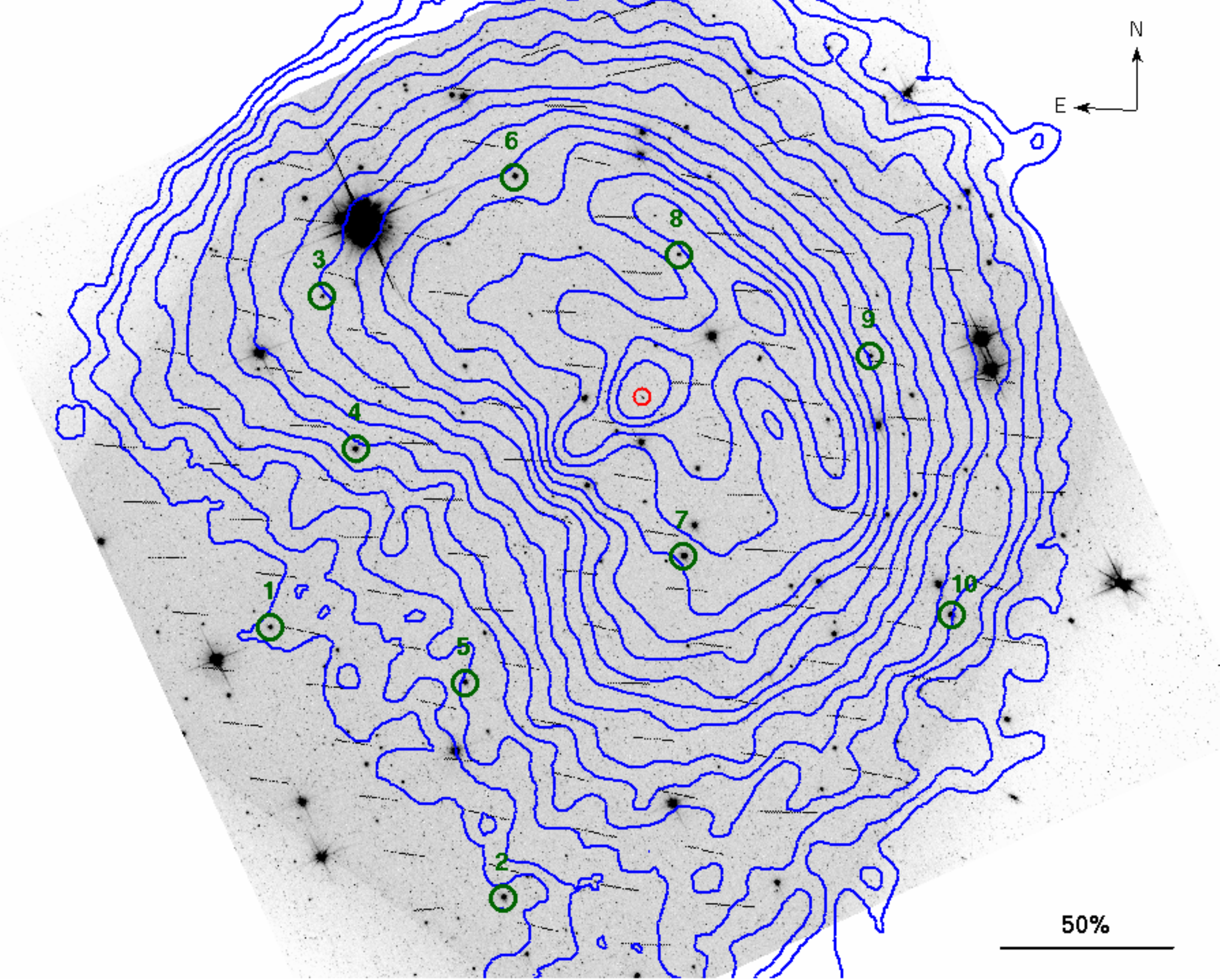}
\caption{
{\em HST} image of the Vela pulsar field taken with the ACS/WFC on 2011 Feb 18 with the F606W filter and the POL0V polariser. The FOV is $\approx$ 102$\times$102 arcsec$^{2}$. The frame has been aligned North-South. The location of the pulsar is marked by the red circle. The reference stars used for analysis are also marked with magenta circles and labelled. The observed polarisation vector map is superimposed in black. The legend shows the vector magnitude for 50\% polarisation. Lastly, the \textit{Chandra}  ACIS (1--8 keV) X-ray contours of the Vela X-ray PWN are overlaid in blue \citep{Pavlov03}.}
\label{figure1}
\end{figure*}

For each visit, the raw images, which had already been flat-fielded, were geometrically aligned, combined and averaged with cosmic-ray removal using IRAF (see Fig. \ref{figure1}). We used a total of five field stars and the IRAF task \textit{ccmap} and the \textit{2MASS} catalogue \citep{Skrutskie06} to fit the ACS/WFC astrometry. The pulsar was found at $\rm \alpha=08^{\rm h} 35 ^{\rm m} 20\fs578\pm0\fs004$, $\rm \delta=-45\degr10\arcmin34\farcs560\pm0\farcs077$ (the errors denote the rms of the astrometric fit). For each set of observations, the images taken in the different polarisers were analysed using the IMPOL\footnote{http://www.stecf.org/software/IRAFtools/stecf-iraf/impol} software \citep{Walsh99}, which produces polarisation maps (see Fig. \ref{figure1}). In order to increase the number of counts, the size of the cells that we used for the mapping is $\approx 0.8 \times 0.8$ arcsec$^2$.


\subsection{Polarisation measurements}

In order to measure its degree of linear polarisation, per each observation we performed aperture photometry of the pulsar in each of the images taken with the  POL0V, POL60V, and POL120V polariser elements using the IRAF task \textit{phot}.  Since one of our goals was to search for possible short-term variability of the pulsar polarisation, we preferred not to co-add all the ten available exposures in each polariser element (Table 1) but to work on each observation at once. Moreover, since they were performed with slightly different spacecraft roll angles (by $\sim\pm1\degr$) and with the pulsar located at slightly different positions on the CCD (by $\sim\pm2$ pixels), this approach allowed us to deal with possible systematics affecting the measurement of the pulsar polarisation and gave us a better handle on the random error estimates. 

As done in \citet{Moran13}, we tested our method by performing aperture photometry on a number of reference stars (see Fig. \ref{figure1}) selected to cover uniformly the ACS/WFC FOV, so as to check for possible systematics, such as the dependence of the linear polarisation on the star position on the CCD. The stars are not saturated but bright enough to provide good statistics and sample different brightness ranges to check for a possible dependence of the polarisation parameters on the star flux. We also wished to verify the size of the aperture to use for polarimetry. We used images taken on 2011 February 18 but with the F550M filter 
and no polariser in place (see Table 1). To maximise the signal--to--noise, we used an aperture of radius 0\farcs2 to measure the flux from the pulsar. We measured the sky counts using an annulus of width $\approx$ 0\farcs1, located 0\farcs15 beyond the central aperture. Then, we applied an aperture correction to our photometry measurement. We then compared the F550M flux of the pulsar with 
that of \citet{Mignani01}, obtained using the {\em HST}/WFPC2 and the F555W filter ($\lambda=550$ nm, $\Delta\lambda=120$ nm), and found consistency once the difference between the two filters are taken into account.


 \begin{table*}
 \caption{P.D. (per cent) of the the Vela pulsar and reference stars as a function of time. The errors for the pulsar are purely statistical, whereas those of the stars are the conservative estimate($\approx$ 1\%) imposed by the debiasing correction \citep{Simmons85}.}
 \label{table1}
 \begin{tabular}{lccccccccccc}
  \hline
  Date & Vela & Star 1 & Star 2 & Star 3 & Star 4 & Star 5 & Star 6 &Star 7 & Star 8 & Star 9 & Star 10\\
  \hline
  2011 Feb 18   &6.6$\pm$0.9   &2.7$\pm$1.0			&1.5$\pm$1.0		&2.0$\pm$1.0	&2.1$\pm$1.0  			&2.7$\pm$1.0  	&1.7$\pm$1.0			&2.2$\pm$1.0	&1.1$\pm$1.0 			&1.8$\pm$1.0	&3.1$\pm$1.0\\
  2011 Feb 19   &10.2$\pm$1.0 &1.4$\pm$1.0  			&1.4$\pm$1.0		&1.6$\pm$1.0	&0.9$^{+1.0}_{-0.9}$	&2.5$\pm$1.0	&0.7$^{+1.0}_{-0.7}$	&2.1$\pm$1.0 	&0.8$^{+1.0}_{-0.8}$	&1.3$\pm$1.0	&1.8$\pm$1.0\\
  2011 Feb 20   &8.2$\pm$0.9   &1.0$\pm$1.0			&0.9$^{+1.0}_{-0.1}$	&1.7$\pm$1.0	&2.3$\pm$1.0			&2.6$\pm$1.0  	&1.4$\pm$1.0			&1.8$\pm$1.0	&2.5$\pm$1.0			&2.3$\pm$1.0 	&2.2$\pm$1.0\\
  2011 Feb 21   &7.2$\pm$0.9   &0.5$^{+1.0}_{-0.5}$	&1.8$\pm$1.0		&3.9$\pm$1.0	&2.1$\pm$1.0  			&1.3$\pm$1.0	&1.4$\pm$1.0			&1.6$\pm$1.0	&1.7$\pm$1.0 			&2.3$\pm$1.0 	&2.8$\pm$1.0\\
  2011 Feb 23   &8.4$\pm$0.9   &0.1$^{+1.0}_{-0.1}$		&1.3$\pm$1.0		&2.7$\pm$1.0	&1.3$\pm$1.0			&1.6$\pm$1.0	&0.7$^{+1.0}_{-0.7}$ 	&1.2$\pm$1.0 	&1.4$\pm$1.0			&1.8$\pm$1.0	&1.4$\pm$1.0\\	
  \hline
 \end{tabular}
\end{table*}

\begin{figure*}
\centering
\subfloat{\includegraphics[width=55mm]{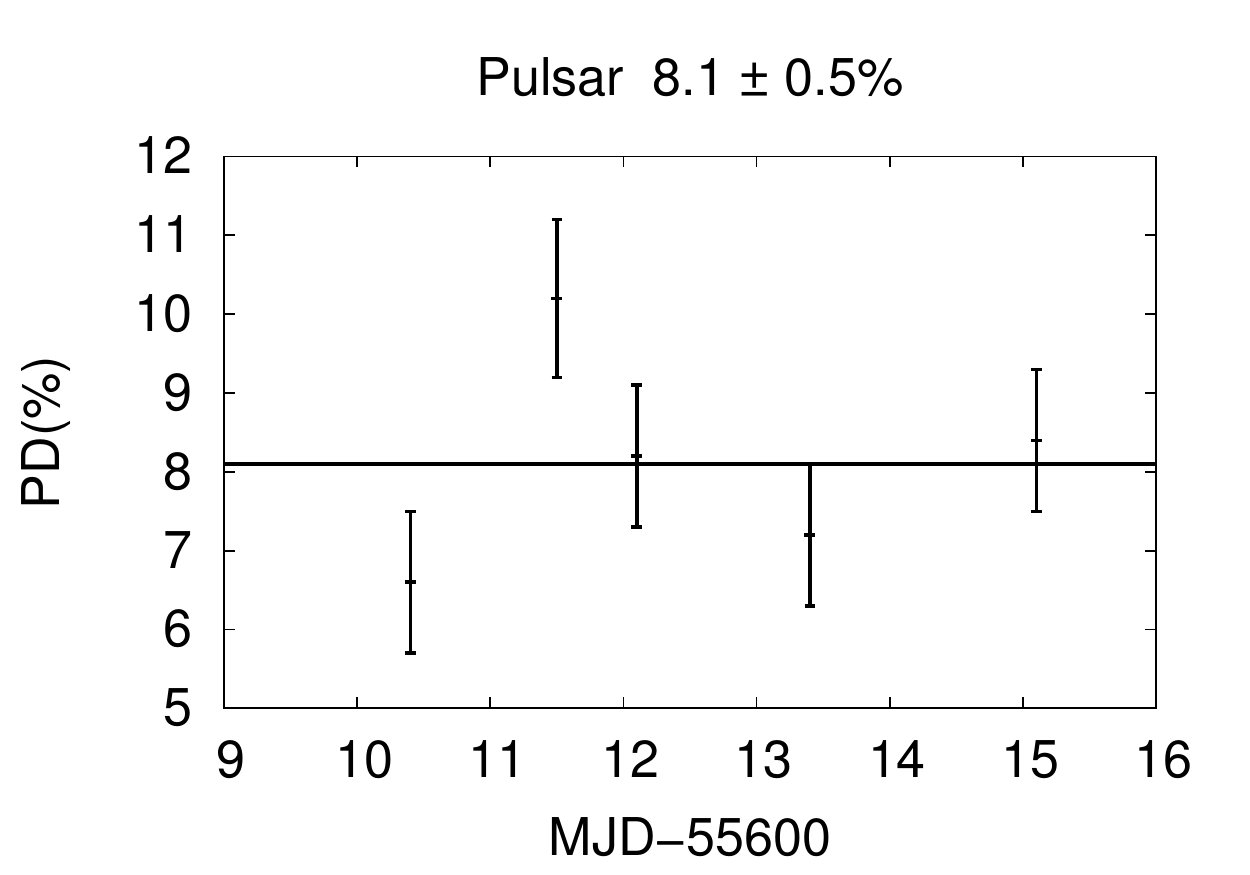}}
\subfloat{\includegraphics[width=55mm]{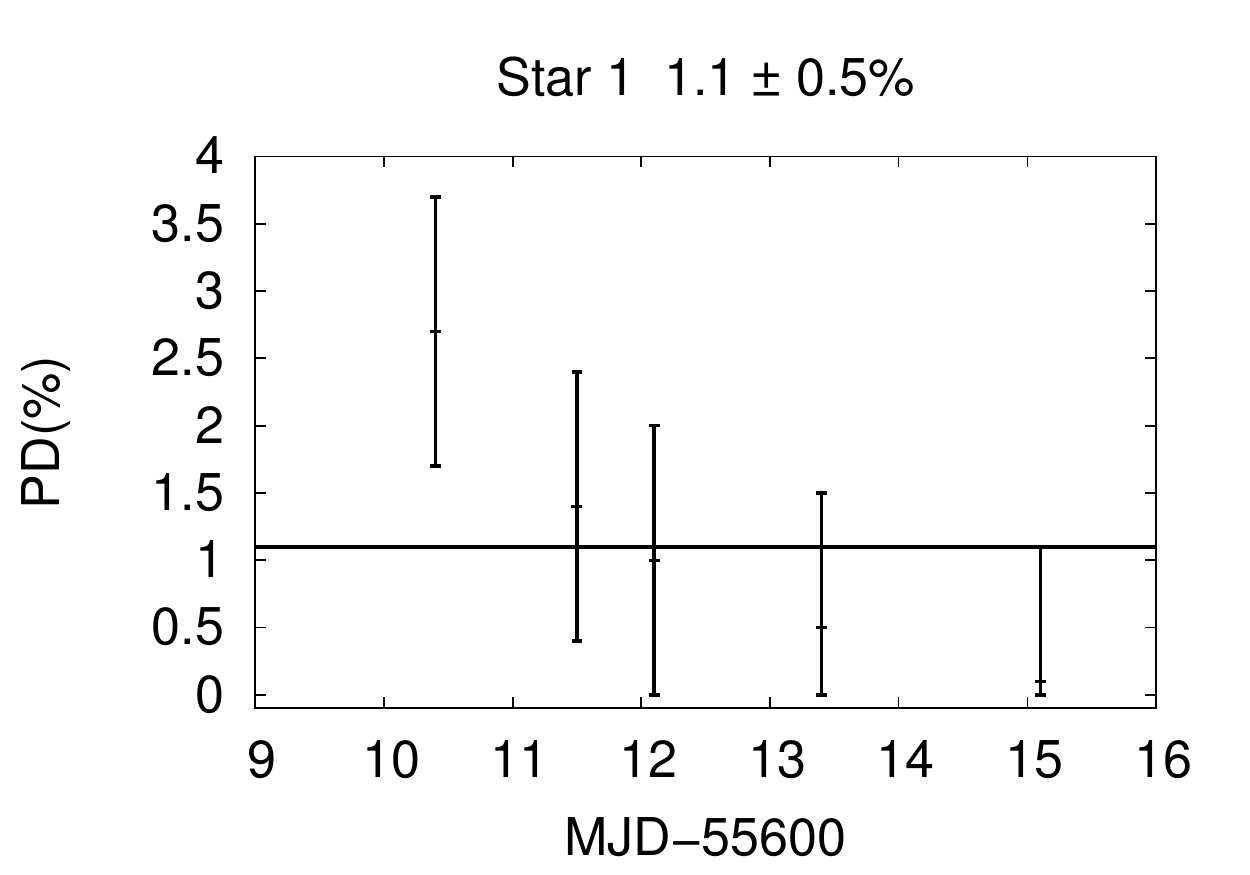}}
\subfloat{\includegraphics[width=55mm]{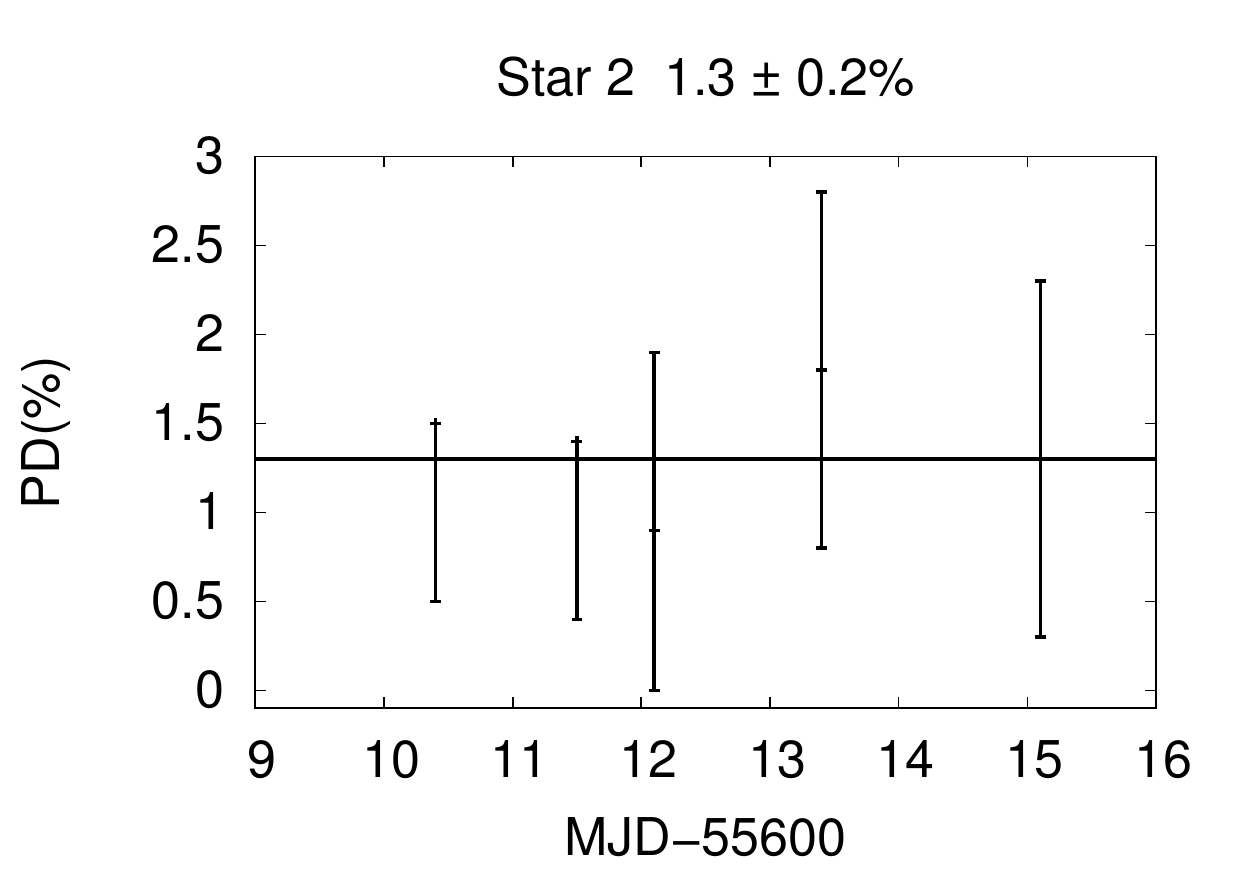}}

\subfloat{\includegraphics[width=55mm]{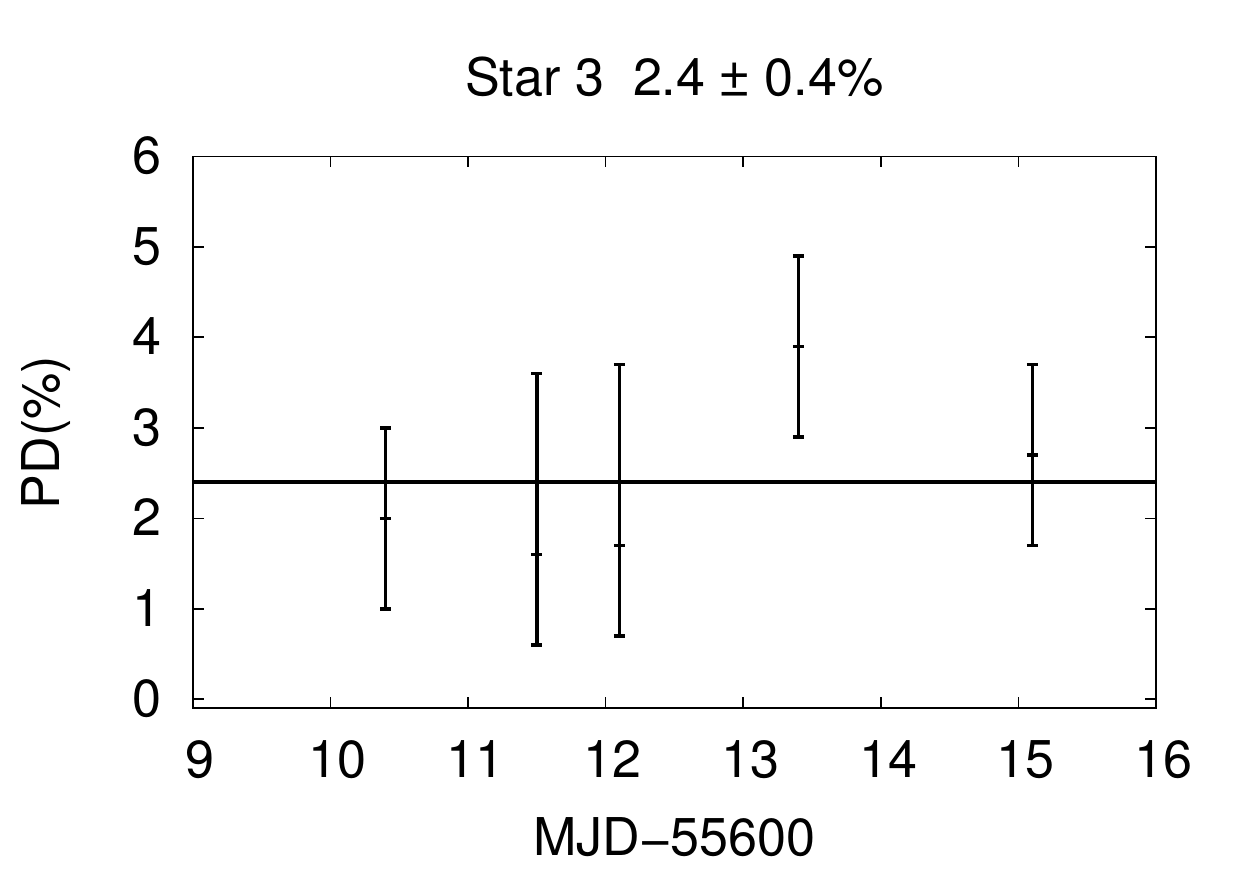}}
\subfloat{\includegraphics[width=55mm]{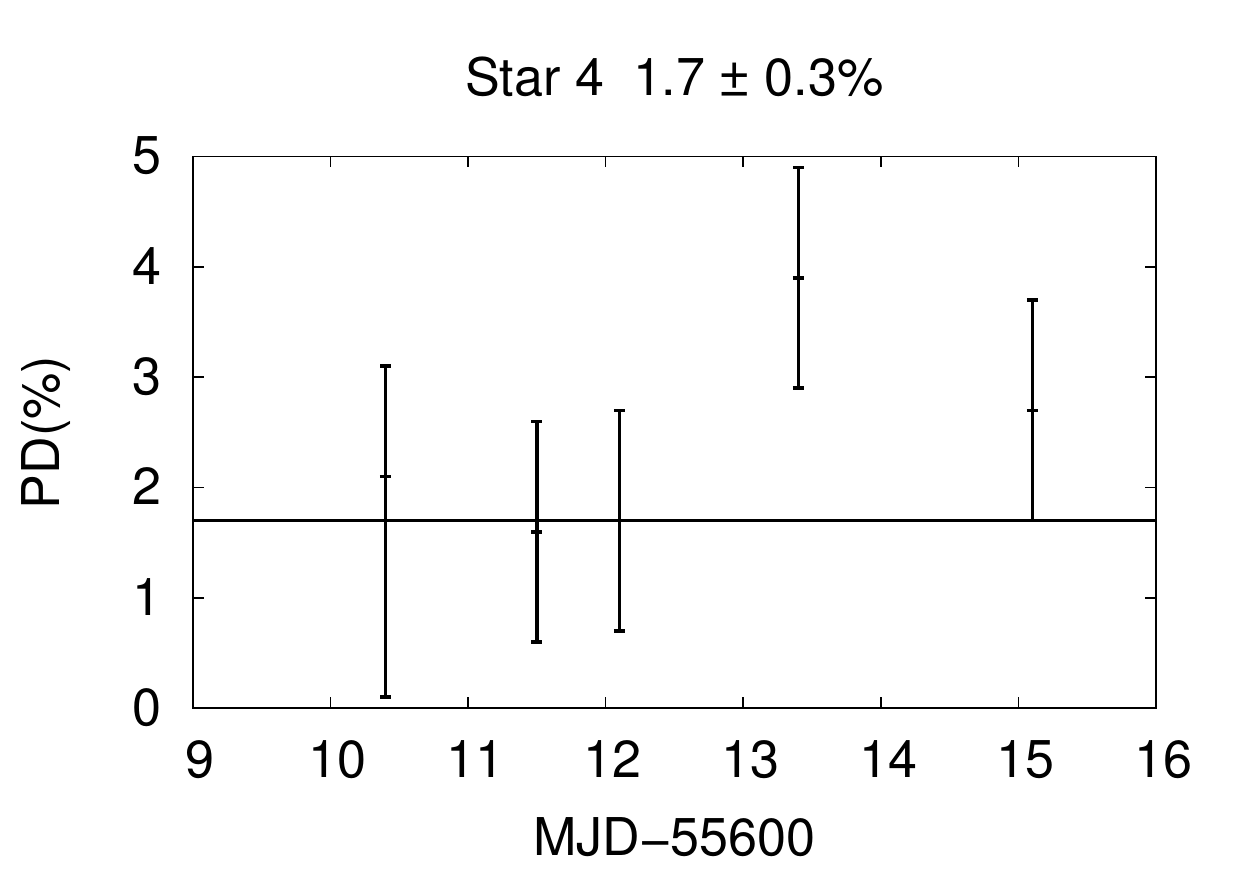}}
\subfloat{\includegraphics[width=55mm]{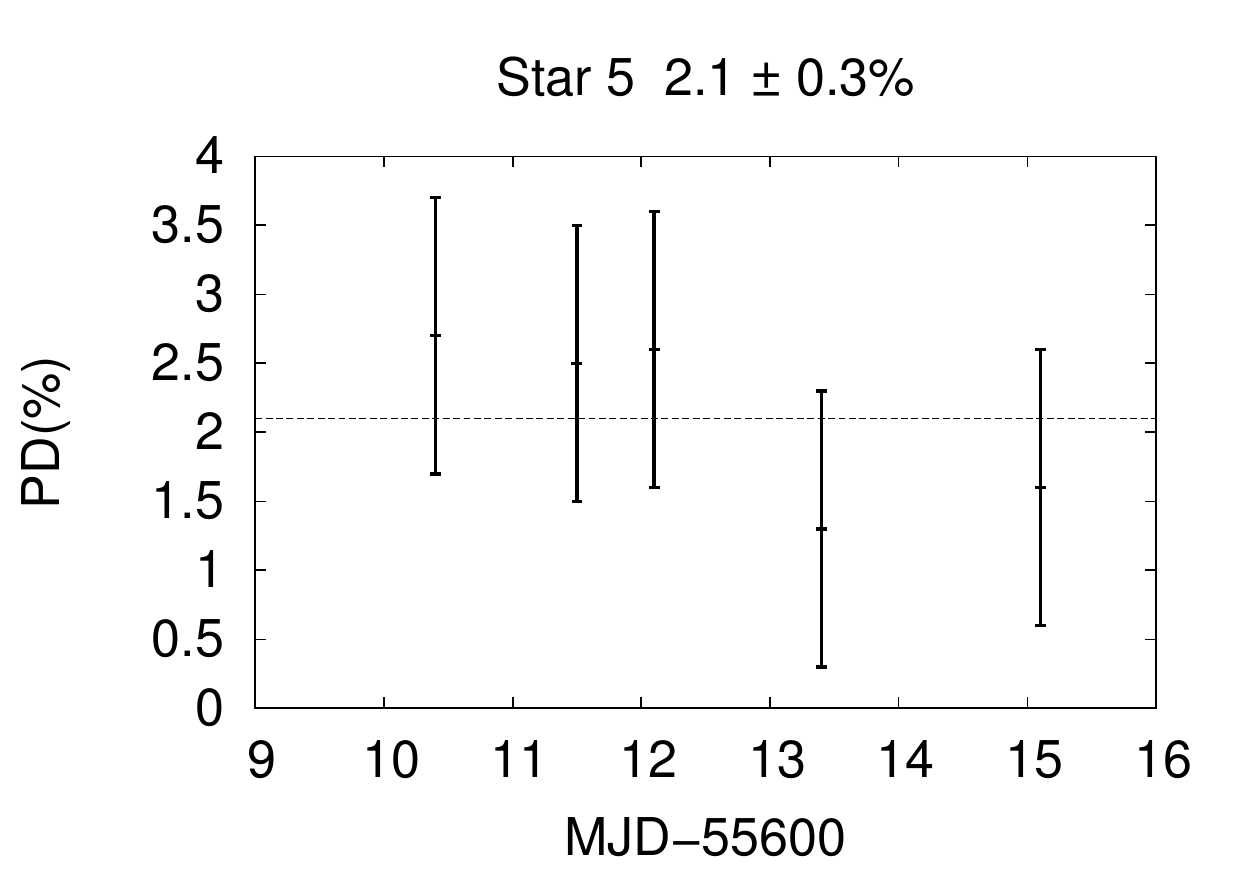}}

\subfloat{\includegraphics[width=55mm]{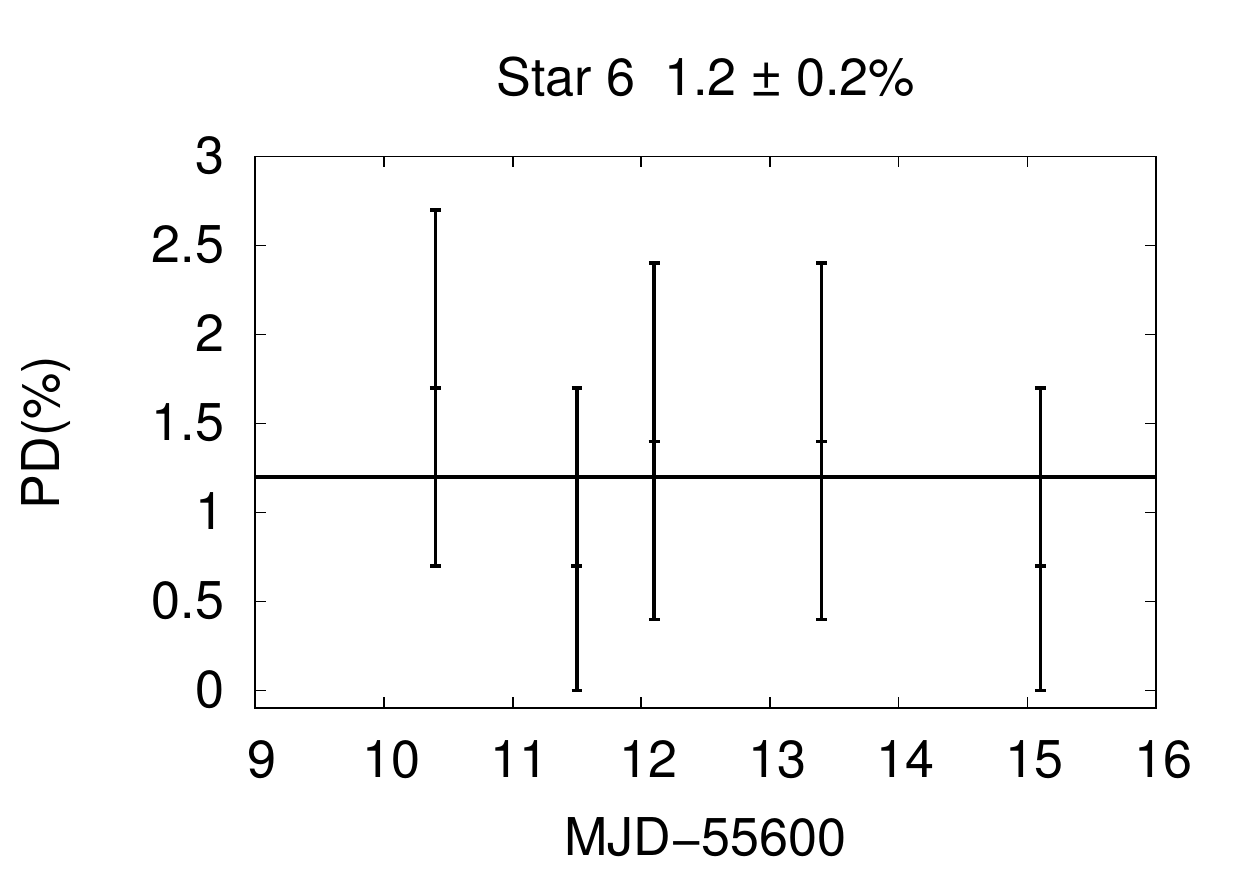}}
\subfloat{\includegraphics[width=55mm]{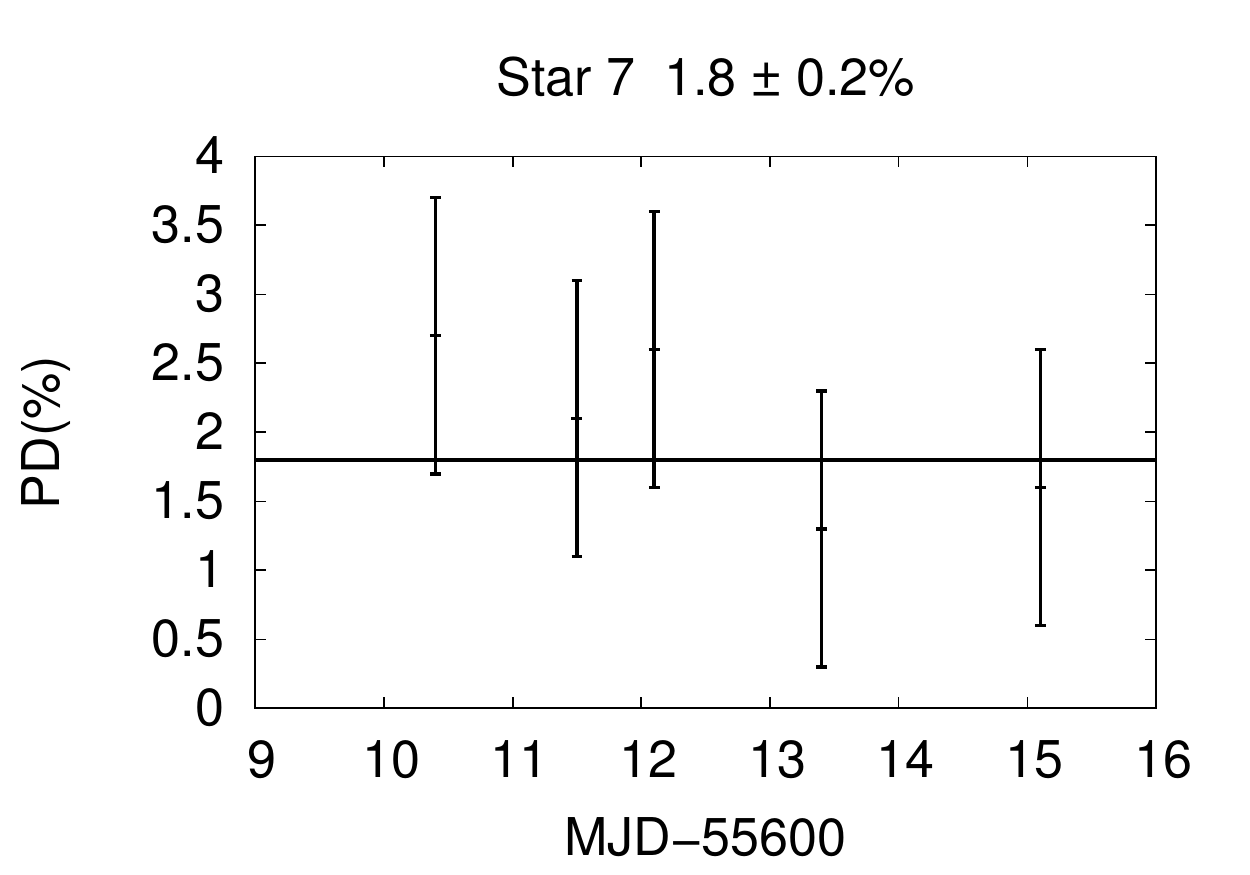}}
\subfloat{\includegraphics[width=55mm]{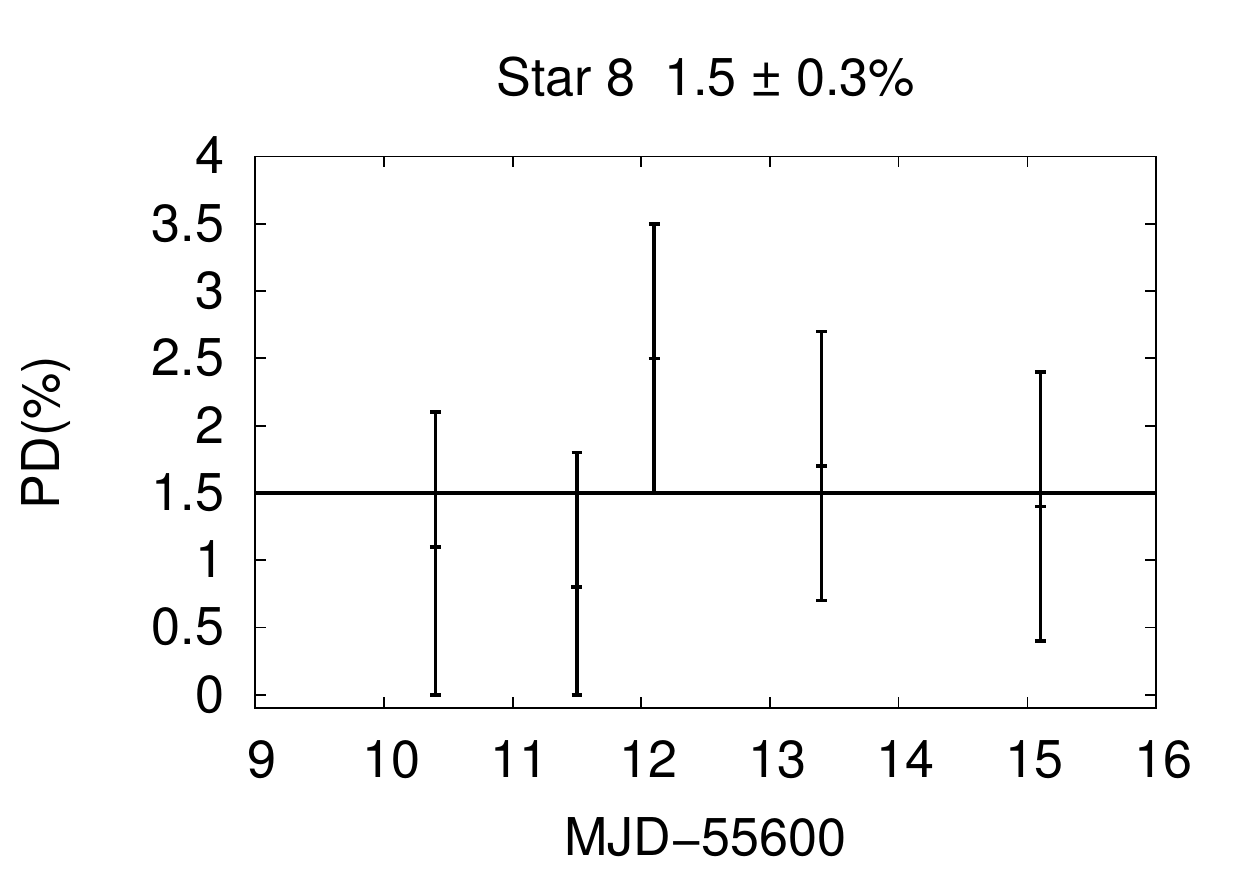}}

\subfloat{\includegraphics[width=55mm]{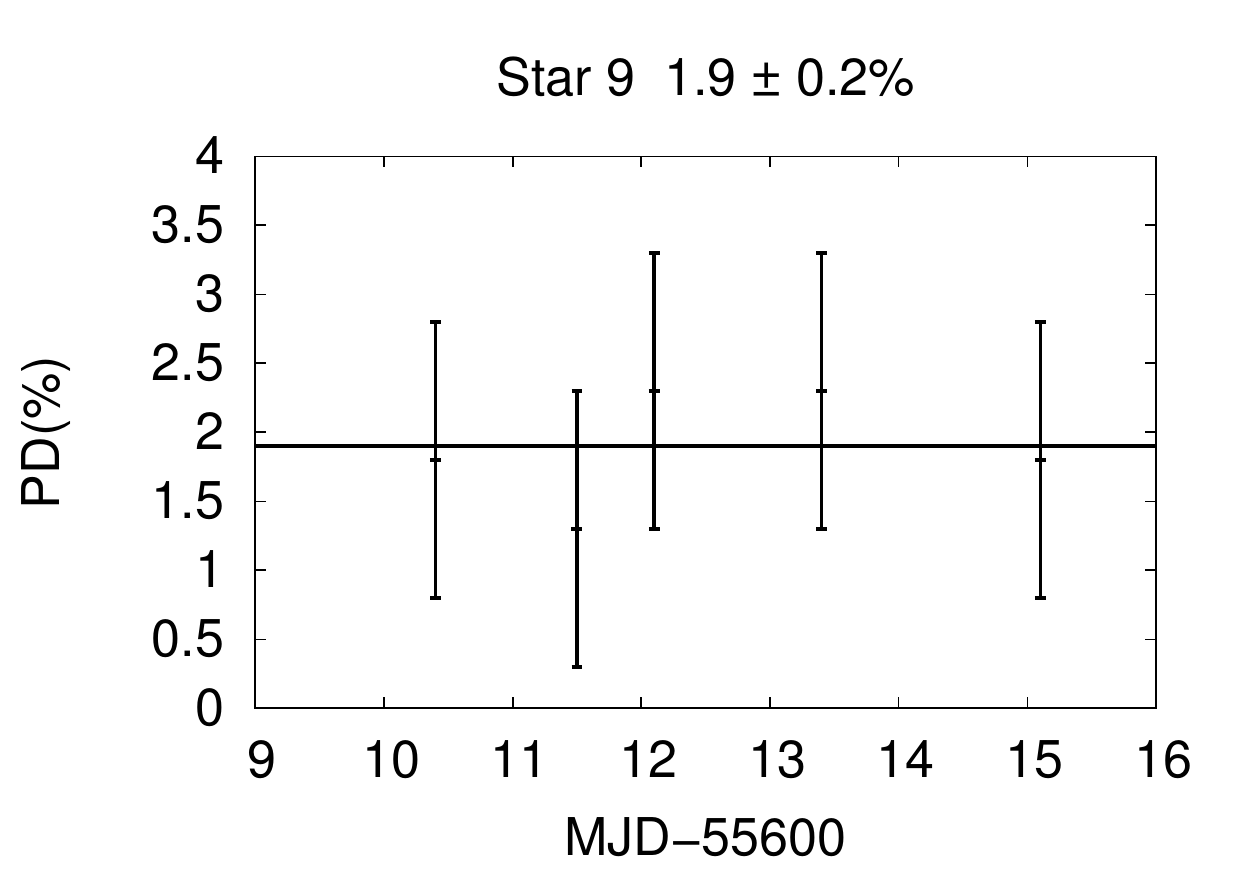}}
\subfloat{\includegraphics[width=55mm]{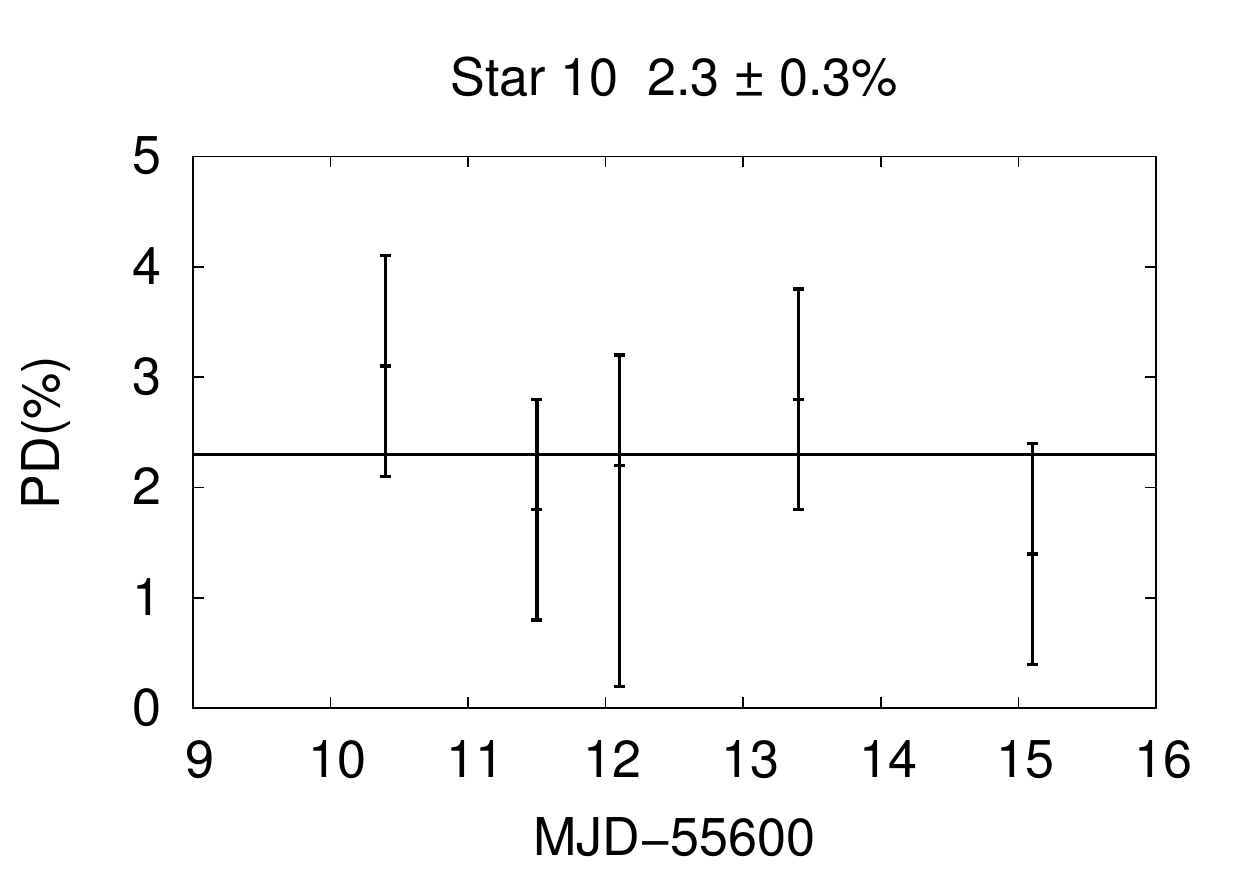}}
\caption{Plots of the P.D. (per cent) of both the Vela pulsar and the reference stars as a function of time. The solid lines are the 
mean of the P.D. Mean values and rms are reported on top of each panel. The errors for the pulsar are purely statistical, whereas those of the stars are the upper limit ($\approx$ 1\%) imposed by the debiasing correction \citep{Simmons85}.}
\label{figure3}
\end{figure*}

To calculate the Stokes parameters, and hence the degree of linear polarisation (P.D.) and position angle (P.A.) we employed the formulae of \citet{Pavlovsky04}  and followed the same approach as in \citet{Moran13}.

\citet{Sparks99} have investigated the achievable accuracies using the assumption of three perfect polarisers oriented at the optimal 60\degr relative position angles (like the ACS/WFC). They found that the important parameter in experiment design is the product of expected fractional polarisation and signal-to-noise (p $\times$ $\langle \rm S/N\rangle_i$). The error on the fractional polarisation, $p$, is just the inverse of the S/N per image. Below are the formulae used for calculating the error in P.D. (Eqn. 1) and P.A. (Eqn. 2) \citep{Pavlovsky04}:

\begin{equation}
\rm log \left(\frac{\sigma_{p}}{p} \right) = -0.102 - 0.9898 \ log \left(p\times  \langle S/N\rangle_i \right)  \\ 
\end{equation}

\begin{equation}
\rm log \ \sigma_{\theta}  = 1.514 - 1.068 \ log \left(p \times \langle S/N\rangle_i \right) \\ 
\end{equation}
\\
where $\langle \rm S/N\rangle_i$ is the average target S/N of the three input images, taken with the three different polarisers.

As a guide to our analysis, we also analysed a number of reference stars selected in the pulsar field (see Fig. \ref{figure1})  to confirm the methodology which we used.

As in the case of the pulsar, we used an aperture of radius 0\farcs2 to measure the flux from each star. The sky counts were measured using an annulus of width $\approx$ 0\farcs1, located 0\farcs15 beyond the central aperture. Aperture correction was applied as for the pulsar.

We also investigated the effects of photometric losses due to charge transfer efficiency (CTE) in the CCDs of the WFC. The ACS team claim that there is no evidence of photometric losses due to CTE for WFC data taken after 2004  (see \citealt{Pavlovsky04}). Nonetheless, we applied the correction for CTE  to our photometry of both the Vela pulsar and the reference stars and found that it does not change the results of the polarimetry. 


\tabcolsep=0.1cm
\begin{table*}
\footnotesize
 \caption{Polarisation P.A.s ($^{\circ}$) of the Vela pulsar and reference stars as a function of time. Reported errors are purely statistical. }
 \label{table2}
 \begin{tabular}{lcccccccccccccc}
  \hline
  Date & Vela & Star 1 & Star 2 & Star 3 & Star 4 & Star 5 & Star 6 &Star 7 & Star 8 & Star 9 & Star 10\\
  \hline
  2011 Feb 18   &142.8$\pm$4.8  	 &175.4$\pm$1.9   	&158.5$\pm$1.7	&131.6$\pm$4.1  	&141.2$\pm$1.2   &142.1$\pm$1.9	&170.6$\pm$1.5	&131.7$\pm$3.5	&124.4$\pm$2.3	&122.9$\pm$4.4	&135.5$\pm$0.8\\
  2011 Feb 19   &152.2$\pm$3.3   	&172.7$\pm$3.8   	&126.8$\pm$1.7	&110.1$\pm$5.1  	&166.9$\pm$2.8   &133.1$\pm$2.0	&144.7$\pm$3.9	&143.7$\pm$1.2	&143.7$\pm$10.5	&126.5$\pm$6.3	&118.4$\pm$1.4\\
  2011 Feb 20   &139.1$\pm$3.8   	&167.9$\pm$5.6   	&160.9$\pm$3.0	&122.1$\pm$4.7  	&145.8$\pm$1.1   &124.0$\pm$2.0	&171.2$\pm$2.4	&119.0$\pm$1.3	&115.0$\pm$3.2	&122.8$\pm$3.4	&138.1$\pm$1.1\\
  2011 Feb 21   &147.9$\pm$4.7   	&80.7$\pm$12.1	&172.5$\pm$1.3	&113.0$\pm$2.0  	&128.1$\pm$1.2   &164.1$\pm$4.1	&176.2$\pm$1.8	&138.6$\pm$1.5	&139.1$\pm$4.8	&126.5$\pm$3.5	&147.7$\pm$0.9\\
  2011 Feb 23   &149.3$\pm$4.0   	&80.8$\pm$44.7   	&119.7$\pm$3.2	&139.1$\pm$2.9  	&92.6$\pm$2.0     &113.3$\pm$3.2	&67.9$\pm$4.8		&113.2$\pm$2.1	&110.7$\pm$5.7	&121.8$\pm$4.6	&132.9$\pm$1.7\\
  \hline
 \end{tabular}
\end{table*} 


\begin{figure*}
\centering
\subfloat{\includegraphics[width=55mm]{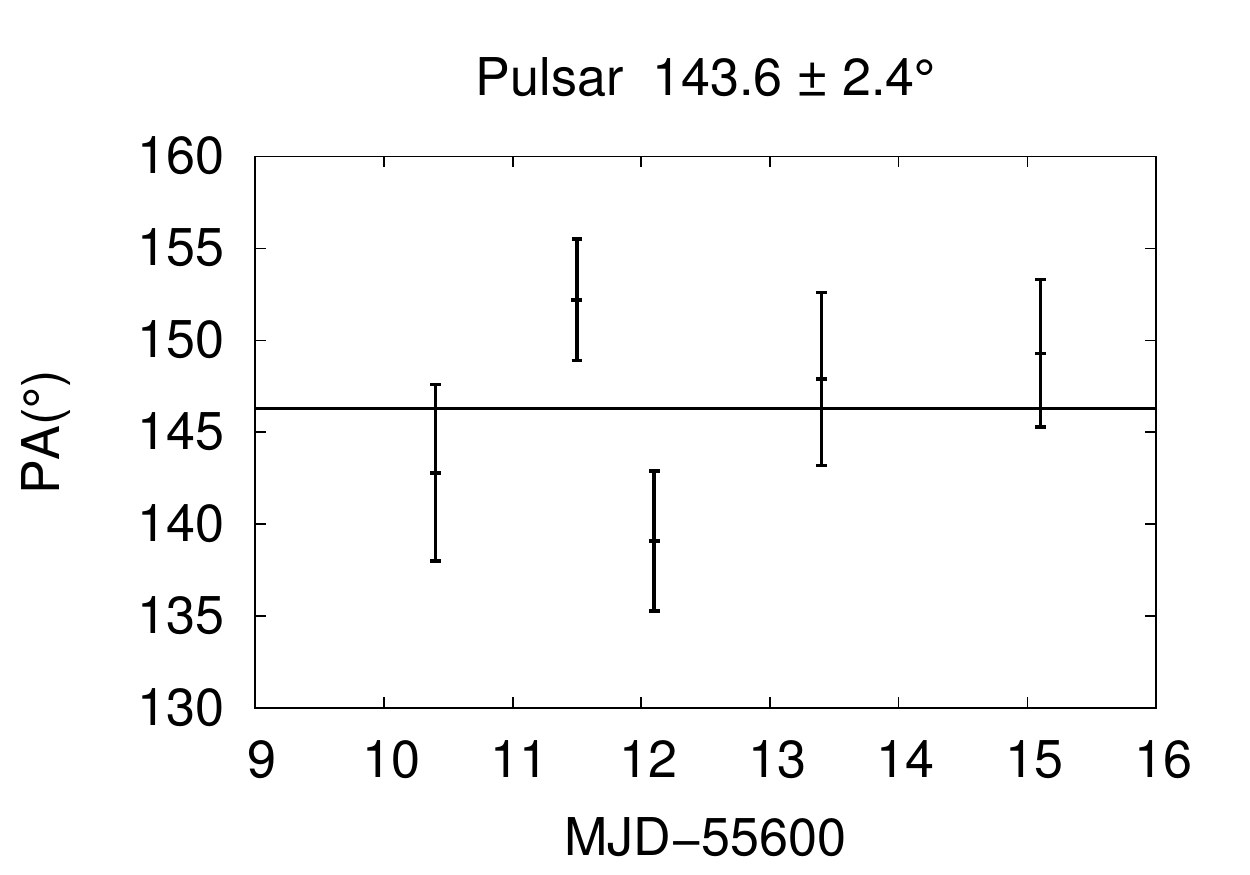}}
\subfloat{\includegraphics[width=55mm]{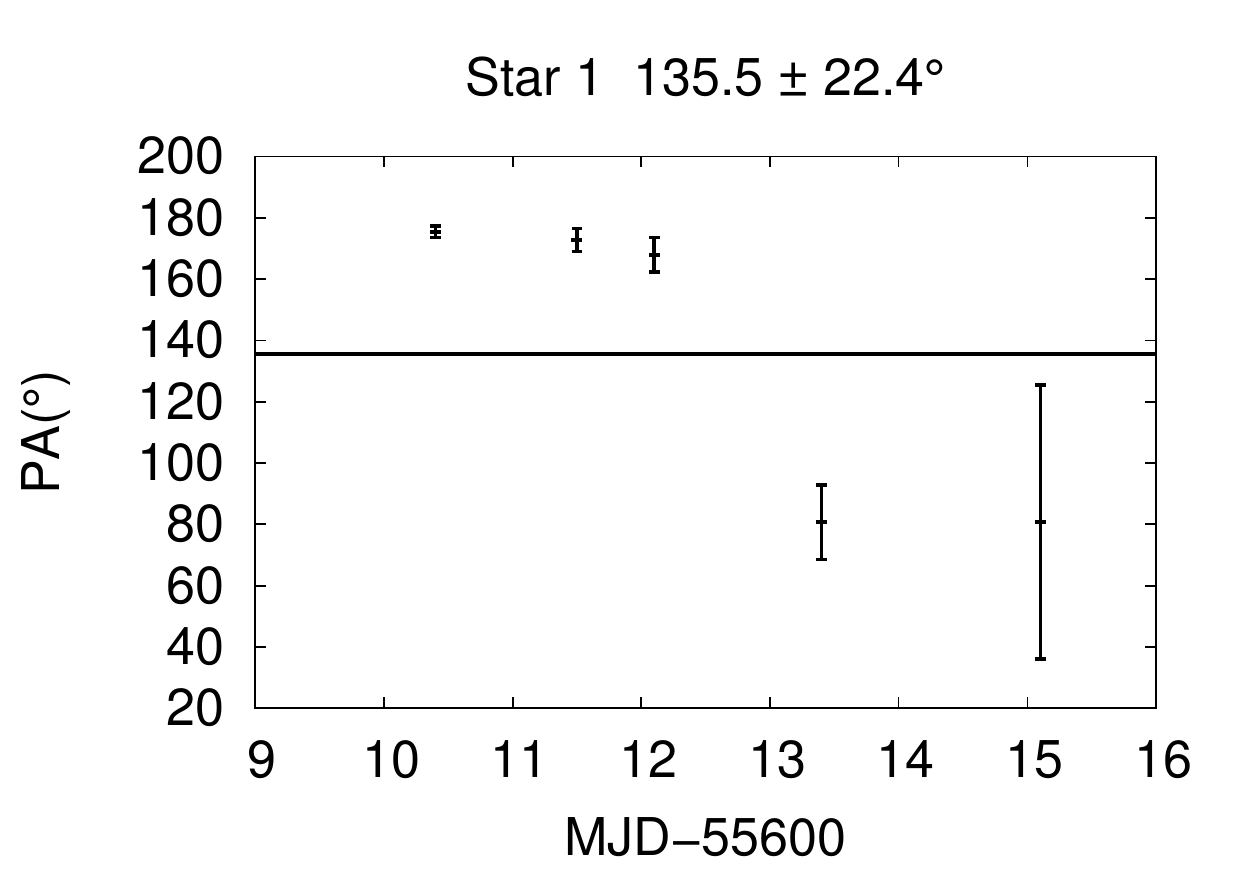}}
\subfloat{\includegraphics[width=55mm]{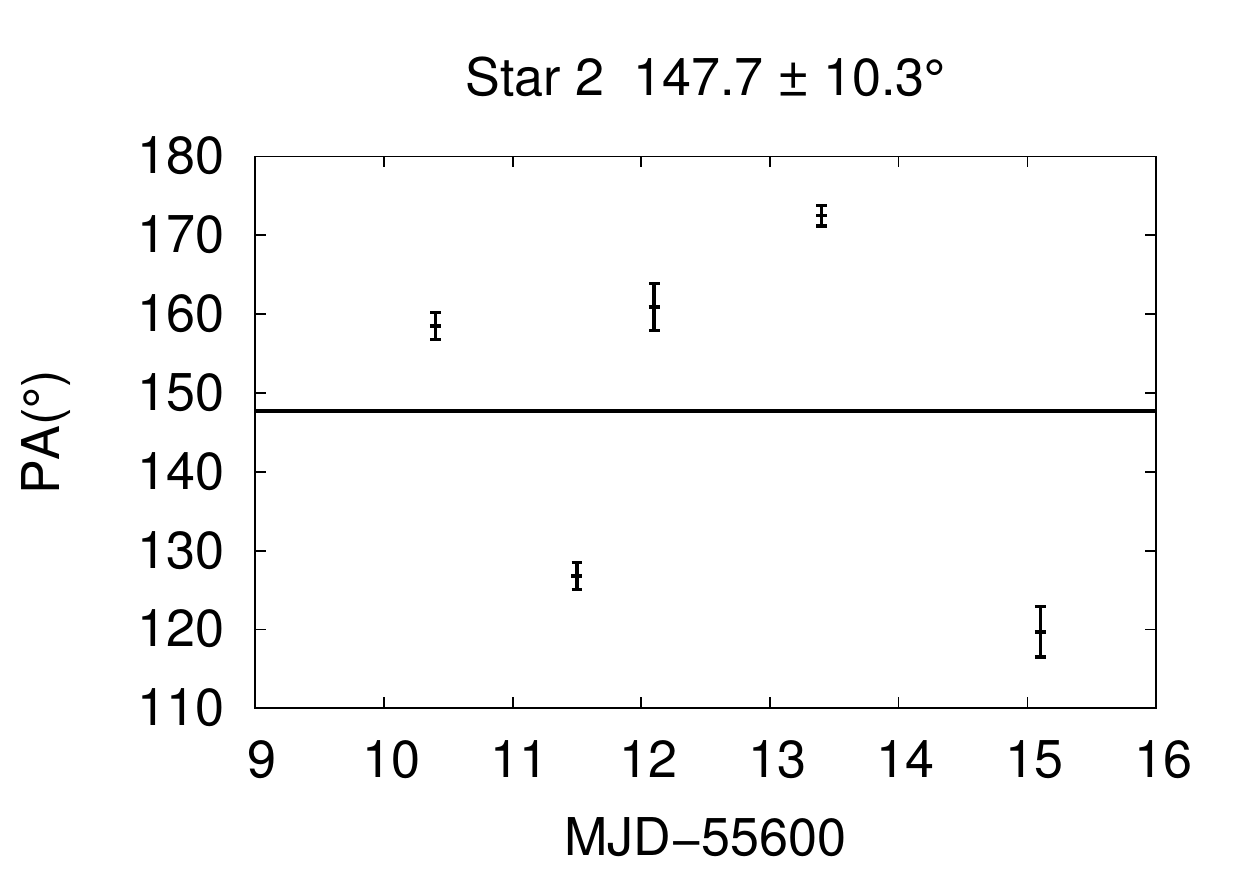}}

\subfloat{\includegraphics[width=55mm]{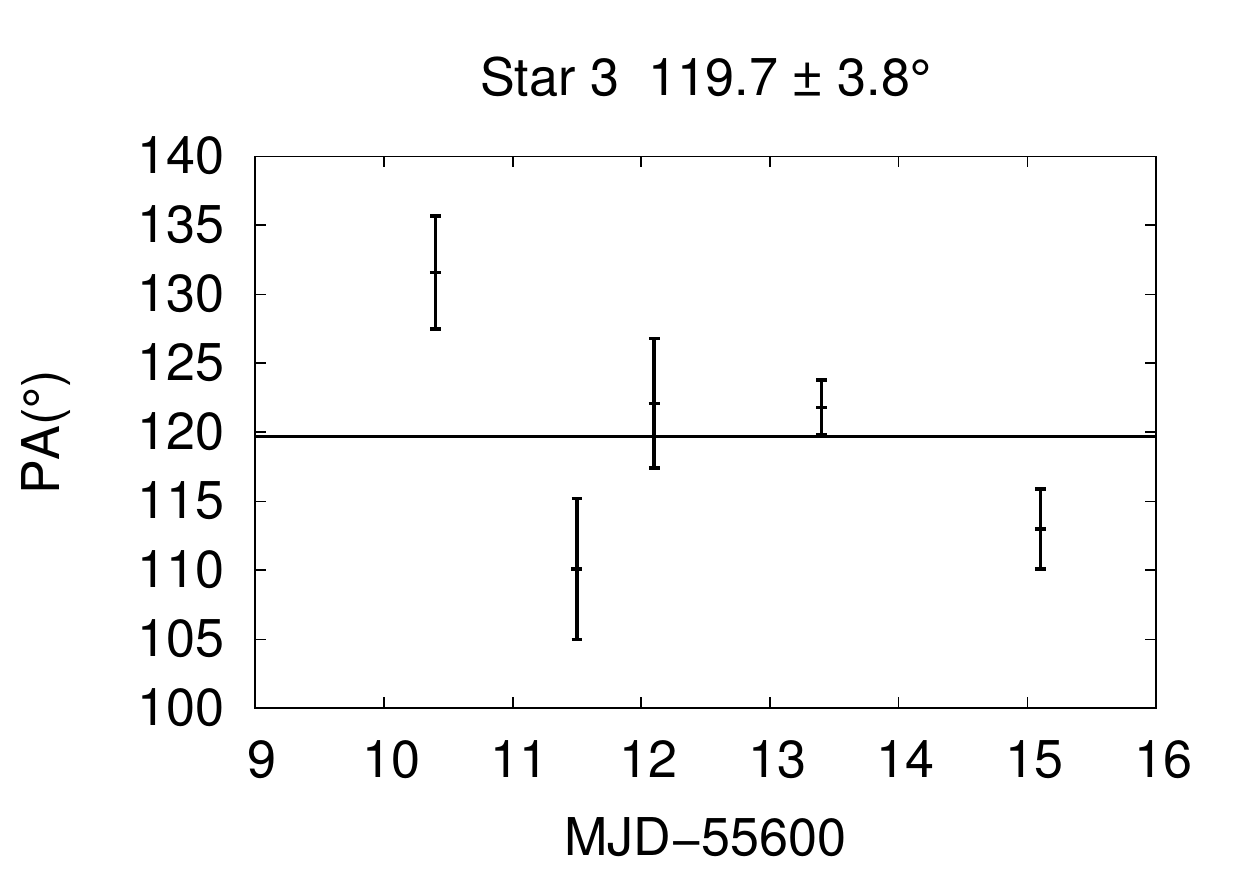}}
\subfloat{\includegraphics[width=55mm]{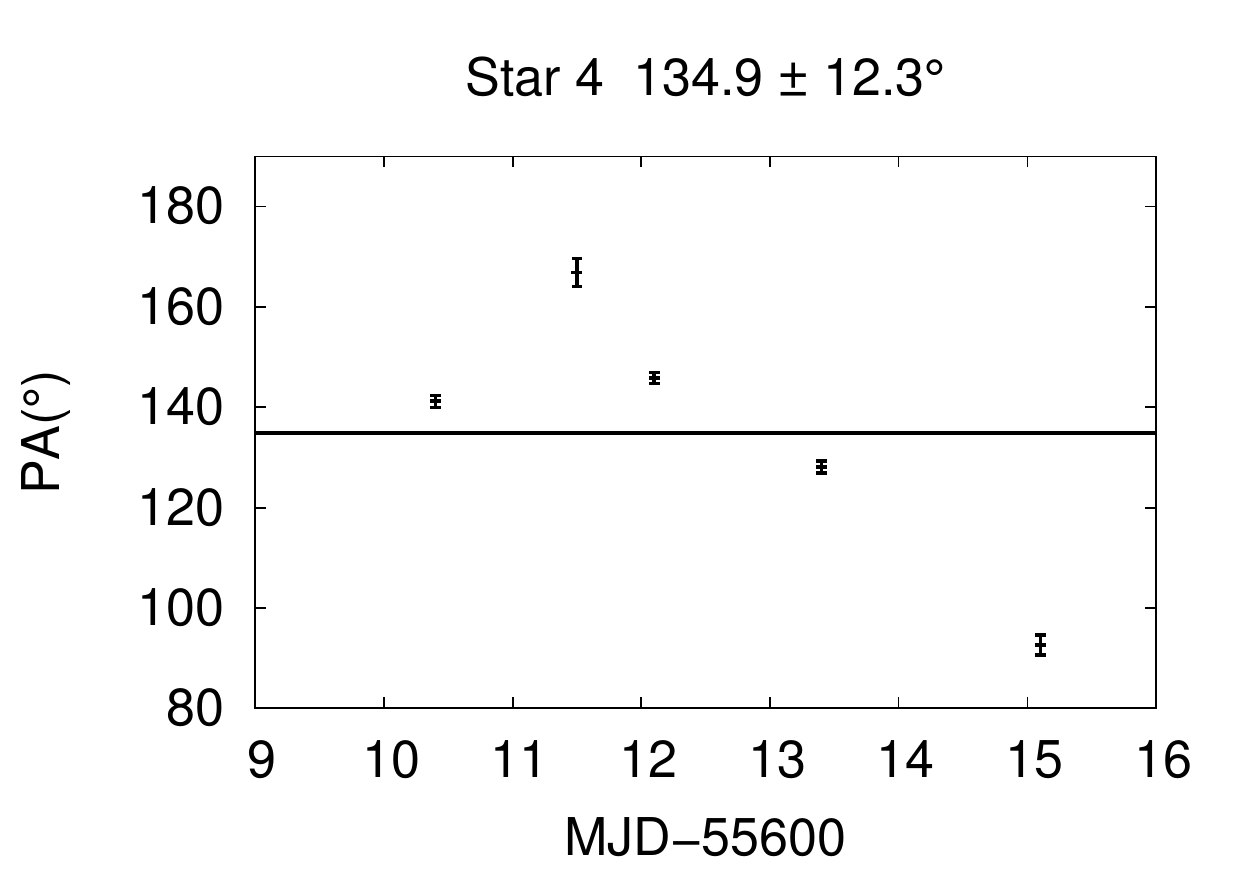}}
\subfloat{\includegraphics[width=55mm]{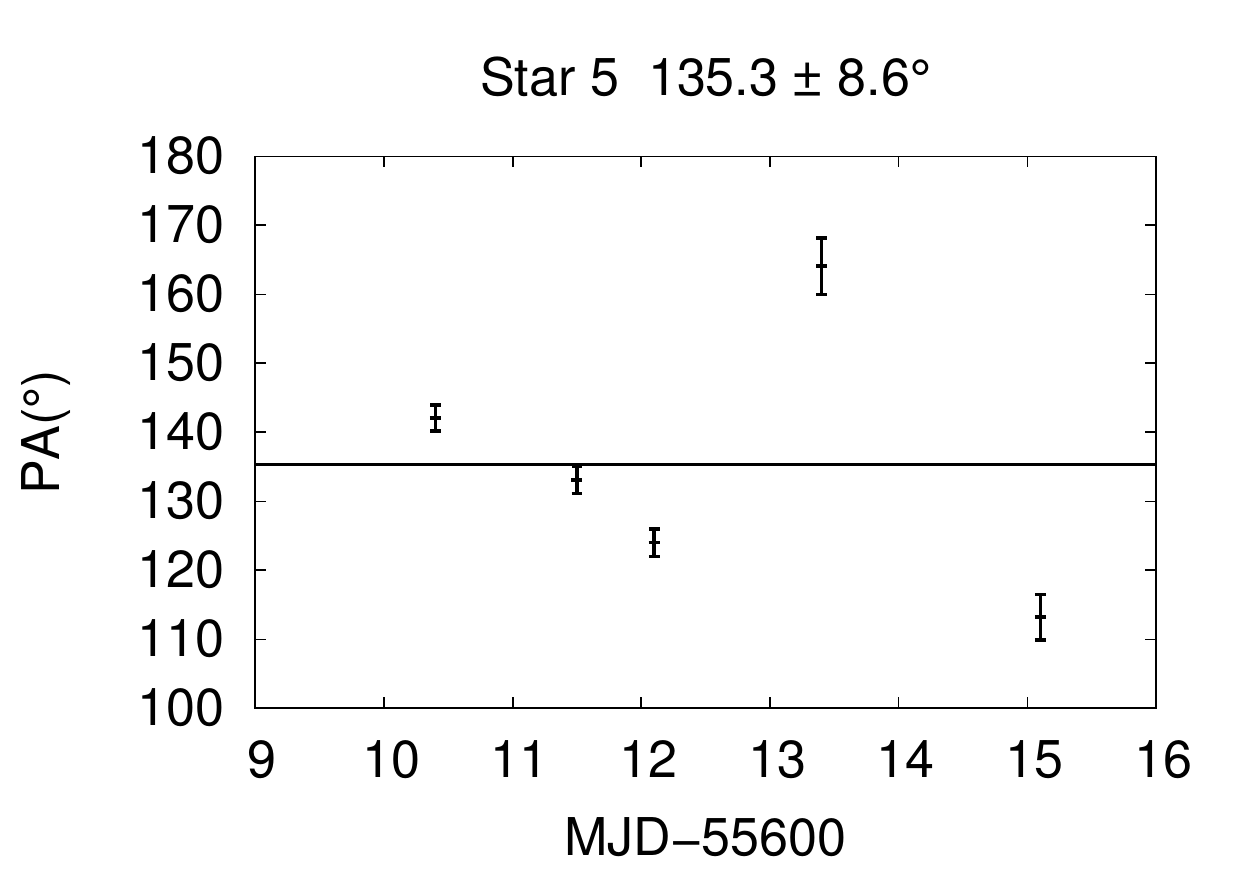}}

\subfloat{\includegraphics[width=55mm]{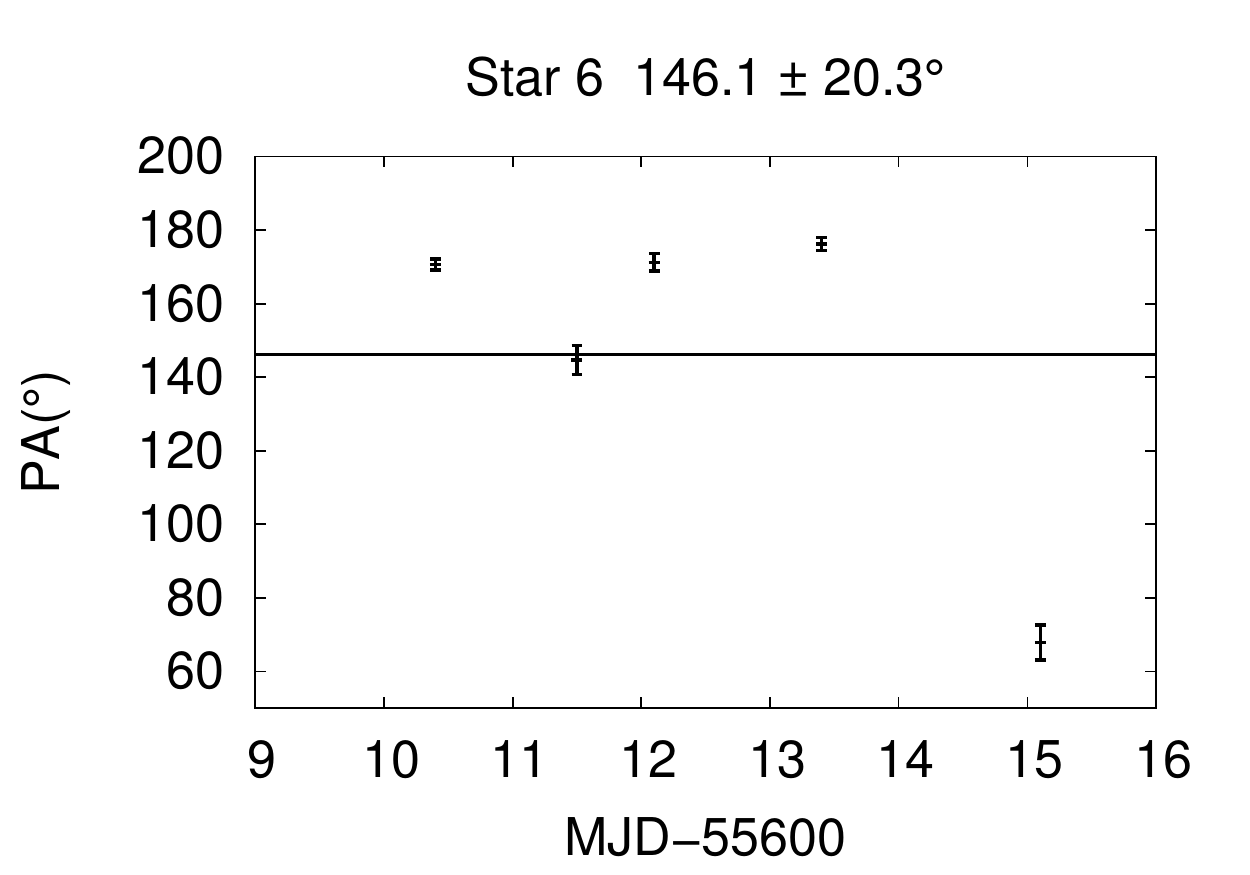}}
\subfloat{\includegraphics[width=55mm]{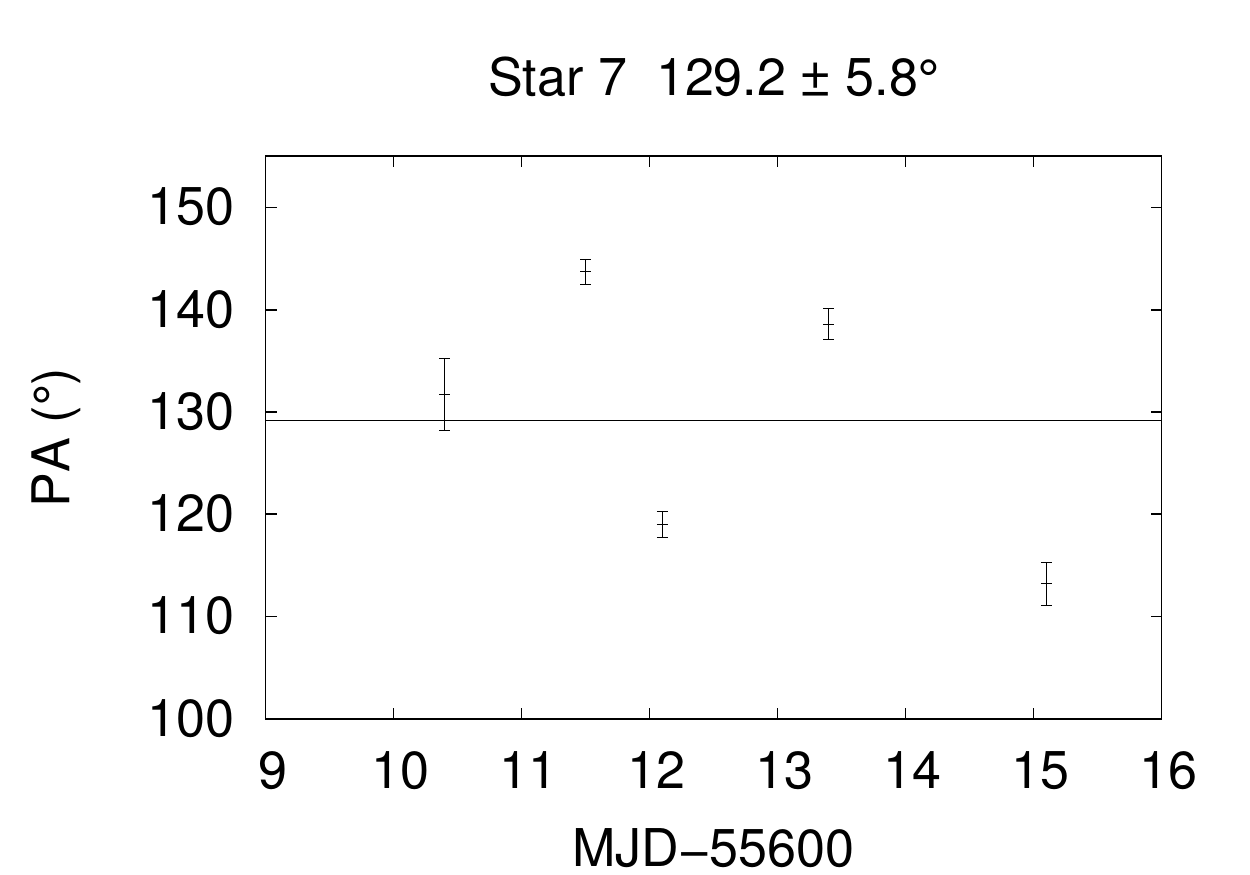}}
\subfloat{\includegraphics[width=55mm]{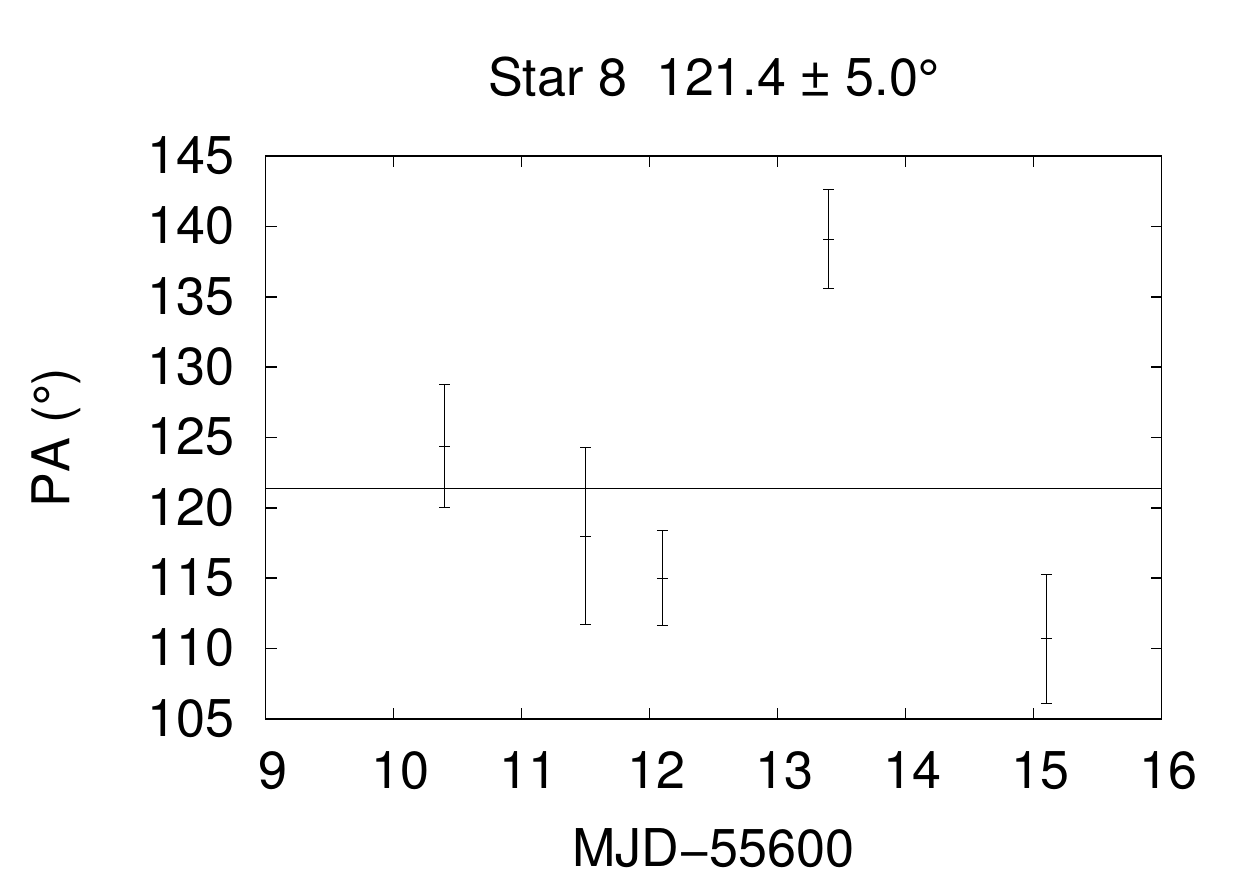}}

\subfloat{\includegraphics[width=55mm]{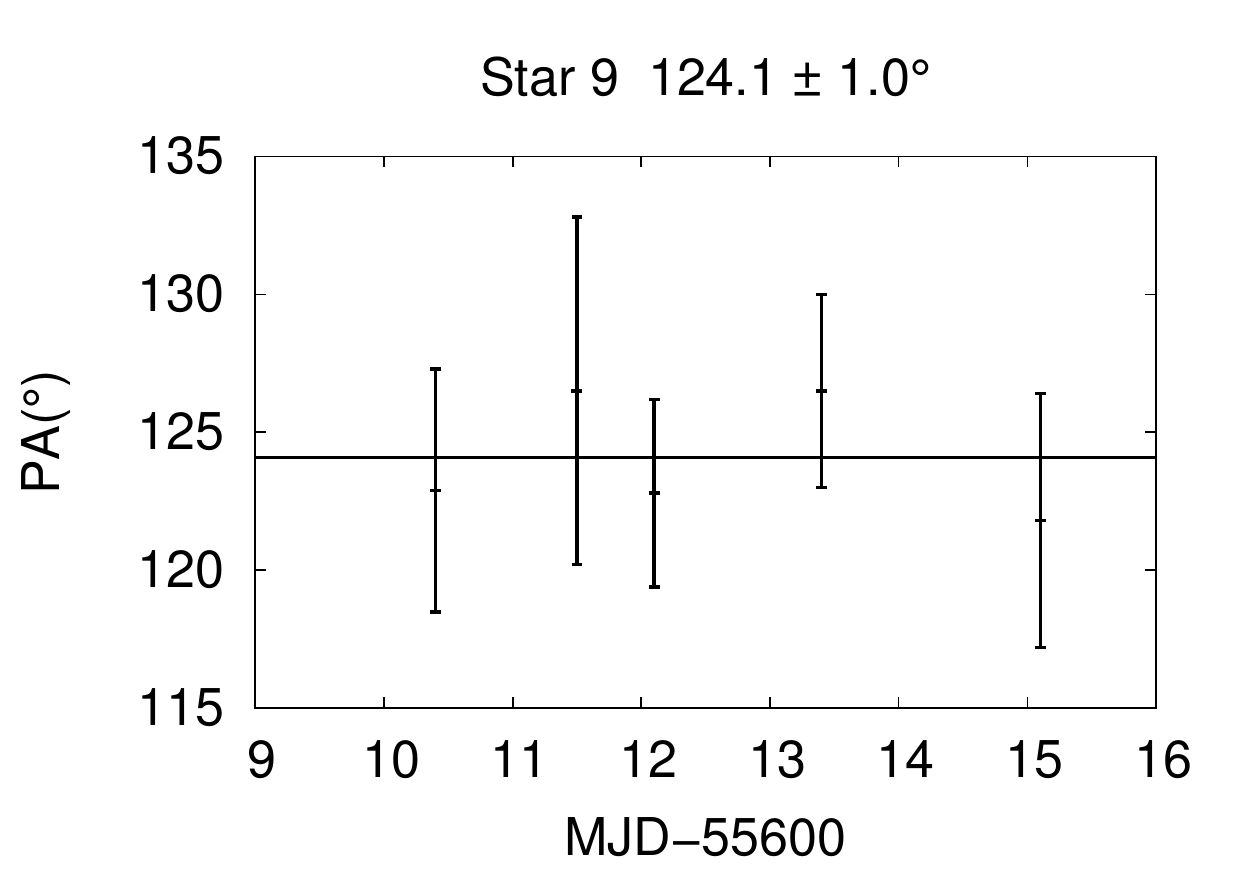}}
\subfloat{\includegraphics[width=55mm]{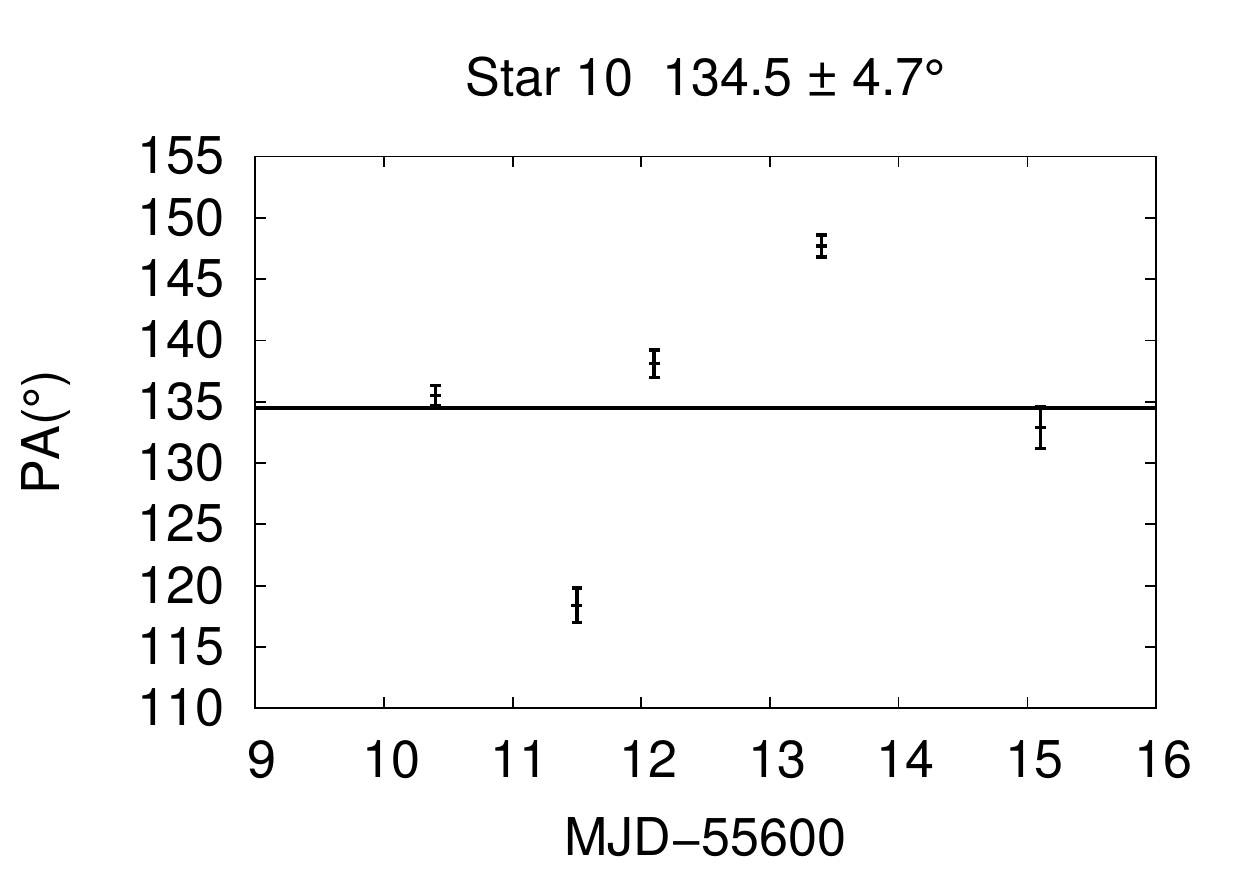}}

\caption{Plots of the P.A. ($\degr$) of both the Vela pulsar and the reference stars as a function of time. The solid lines are the 
mean of the P.A. Mean values and rms are reported on top of each panel.}
\label{figure4}
\end{figure*}


 An important property of polarisation that needs to be considered during our analysis is that of bias. This is due to instrumental errors which tend to increase the observed polarisation of a target with respect to its true polarisation. The effect is negligible when $\eta = p \times S/N$ is high ($> 10$), where $p$ is the fractional polarisation of the target, and S/N is the target signal--to--noise per image. See for example Fig. 4 of \citet{Sparks99}. Since the Vela pulsar is in the high $\eta$ regime, the debiasing correction is small and therefore we omit it. However this is not the case for the reference stars which have low measured polarisation ($<$ 3\% and mostly $<$ 2\%) consequently the associated errors should be $\approx 1\%$ \citep{Simmons85} and can be considered to be zero, and hence have low $\eta$. See section 3.1.

\section{Results}

In this section we present the measurements of P.D. and P.A. for both the Vela pulsar and each of the reference stars in {Fig. \ref{figure1}, obtained per each of the five observing epochs. The values of P.D. and P.A., together with their associated errors, are summarised in Tables 2--3 inclusive. For an easier visualisation of the results, and to make a trend analysis easier, we have also plotted the values of P.D. and P.A. for both the Vela pulsar and each of the reference stars as a function of time (see Figs.\ 2 and 3). 


\subsection{The Pulsar}

As seen from Fig.\ 1, the Vela pulsar is clearly detected in polarised light, both in the POL0V, as well as with the other polariser elements. The P.D. values of the Vela pulsar measured at the different epochs vary from a  minimum of 6.6\% to a maximum of 10.2\%, the latter corresponding to the second observation. The differences between these values are not statistically significant being, in most cases, within $3 \sigma$. The same is true for the P.A. values. We have computed mean P.D. and P.A. values for the Vela pulsar using the values listed in Tables 2 and 3. Using a $\chi^{2}$ goodness-of-fit, we also found no significant variation (at the 95\% confidence level) in the polarisation of the Vela pulsar during the short time span (5 days) covered by these observations. However, more detailed follow-up observations covering a much longer time span will be needed to determine if there is any longer term variation. The results for the Vela pulsar are $\rm P.D.=8.1\%\pm0.6\%$, and $\rm P.A.=146.3\degr\pm2.4\degr$. The values are reported on top of the Vela pulsar panel in Figs.\ 2 and 3 and in Table 4. Not only are these results in good agreement with those of \citet{Wagner00} and \citet{Mignani07}, both obtained using the VLT and the same data set, but they are of greater precision owing to the longer integration time and better spatial resolution of the {\em HST} observations, which allowed us to minimise the contamination from the sky background.

As done for the pulsar,  we present the mean P.D. and P.A. values for all the reference stars (see Table 4). 


\subsection{The Nebula}

The Vela PWN is not detected in polarised optical light in any of the single observations, neither with the POL0V (Fig. \ref{figure1}) nor with any of the other polariser elements (POL60V, POL120V), with the polarised sky background apparently looking uniform. The ACS/WFC polarisation map (Fig. \ref{figure1}) shows the variation of the polarisation vectors in the pulsar field and in the vicinity of the pulsar itself. Each vector has magnitude equal to the degree of polarisation computed over cells of size $\approx 0.8 \times 0.8$ arcsec$^2$ and its orientation corresponds to the position angle measured at that point. Such a map allows one to visualise the direction of the magnetic field lines within the region surrounding the pulsar. From {Fig. \ref{figure1} one can see that the pulsar field is significantly polarised, at the level of $\rm P.D. \approx 10\%$ per cell. However, we cannot see any obvious variation, neither in intensity nor in direction, in the polarisation vectors in the vicinity of the pulsar. This is consistent with the fact that the PWN is not detected in polarised light. This is clearly demonstrated in Fig. \ref{figure1}, where we display the ACS/WFC polarisation map with superimposed X-ray contours of the PWN in the 1--8 keV energy range  as observed by the {\em Chandra}-ACIS detector  \citep{Pavlov03}.  The figure clearly shows that there is no significant change in the polarisation vectors along the main structures of the X-ray PWN, such as the inner and outer arcs and the jets southeast of the pulsar.  Unfortunately, owing to the smaller FOV of the ACS/WFC with the polariser optics in, and the chosen observing strategy, with the Vela pulsar at the centre of the detector, some parts of the region covered by the spatial extent of the X-ray PWN are not fully included in the image or are affected by vignetting at the edge of the detector. In particular, it does not include the region of the bright and long jet protruding northwest of the pulsar position.  Thus, we cannot say anything about the jet polarisation and whether it is variable, like its X-ray flux and morphology \citep{Durant13}. 

In order to increase our sensitivity to the detection of the PWN in polarised light, we co-added all the 30 available exposures taken in each of the three polarisers (Table 1), corresponding to a total integration time of $\approx 40725$ s per element. Since the X-ray PWN is extended over scales of tens of arcsec$^2$, the possible systematics that might have affected the measurement of the pulsar polarisation, such as the slightly different spacecraft roll angle and the pulsar centring on the CCD (see Sectn.\ 3.1),  are much less important here. After co-adding all these exposures, subtracting the field stars, and smoothing with a median filter of 1$^{\prime\prime}$, we found no evidence of the PWN. At the same time, we found no significant difference in the magnitude and orientation of the polarisation vectors of the sky background, meaning that its polarisation properties do not change on time scales as short as a few days, as expected.

We use the value of the sky to determine the upper limit of the optical surface brightness of the Vela PWN. For this measurement, we first used a total of 200 cells, each of which was of 1 arcsec$^2$ area, and evenly spread across the area coinciding with the central part of the  the X-ray PWN and well away from field stars. After measuring the flux of the sky in each cell, we took the mean of these 200 measurements, which yielded a value of 23.6 magnitudes arcsec$^{-2}$. We then used this mean in conjunction with the standard deviation of the flux measurements to determine the upper limit of the surface brightness of the Vela PWN. The 3$\sigma$ upper limit of the PWN is 26.8 magnitudes arcsec$^{-2}$. Due to the limited FOV of these observations (see Fig. 1) there is no means of determining a background outside the area covered by the PWN. Hence, we note that this is quite a conservative upper limit.


\section{Discussion}

\begin{table}
\begin{center}
 \caption{Overall results for the P.D. (per cent) and P.A. ($^{\circ}$). The values listed in the table are the 
 mean and standard error of the P.D. and P.A. The last columns give the star magnitude in the F550M filter ($m_{F550M}$) and the S/N. Photometry errors are purely statistical. }
 \begin{tabular}{lcccc}
  \hline
   & P. D. (\%) & P. A. ($^{\circ}$)   & $m_{F550M}$ & S/N\\
  \hline
  Pulsar		&8.1$\pm$0.6   			&146.3$\pm$2.4	&23.604$\pm$0.018		&60\\
  Star 1		&1.1$\pm$0.5  				&135.5$\pm$22.4	&20.430$\pm$0.004		&543\\
  Star 2		&1.3$\pm$0.2   			&147.7$\pm$10.3	&19.600$\pm$0.004		&1085\\
  Star 3		&2.4$\pm$0.4   			&119.7$\pm$3.8	&21.034$\pm$0.005		&362\\
  Star 4		&1.7$\pm$0.3   			&134.9$\pm$12.3	&19.520$\pm$0.004		&1085\\
  Star 5		&2.1$\pm$0.3   			&135.3$\pm$8.6	&19.903$\pm$0.004		&1085\\
  Star 6		&1.2$\pm$0.2  				&146.1$\pm$20.3	&19.824$\pm$0.004		&543\\
  Star 7		&1.8$\pm$0.2   			&129.2$\pm$5.8	&18.984$\pm$0.004		&1085\\
  Star 8		&1.5$\pm$0.3   			&121.4$\pm$5.0	&21.446$\pm$0.006		&271\\
  Star 9		&1.9$\pm$0.2   			&124.1$\pm$1.0	&21.374$\pm$0.005		&271\\
  Star 10		&2.3$\pm$0.3  				&134.5$\pm$4.7	&19.249$\pm$0.004		&1085\\
  \hline
 \end{tabular}
 \end{center}
\end{table}



We have studied the phase-averaged polarisation properties of the Vela pulsar using archival {\em HST}/ACS data. We have produced polarisation vector maps of the $\approx102\times102 \rm \ arcsec^{2}$ region directly surrounding the pulsar, covering most of the spatial extent of the X-ray PWN,  and measured the degree of linear polarisation and the position angle of the pulsar's integrated pulse beam. This work marks the first high-spatial resolution multi-epoch study of the variability of the polarisation of the pulsar.

The results for the Vela pulsar are $\rm P.D.=8.1\% \pm0.6\%$, and $\rm P.A.=146.3\degr\pm2.4\degr$ (see Table 4). These values firmly confirm those of \citet{Wagner00} and \citet{Mignani07}, both obtained using the same VLT data set but reported with different error estimates (see \citet{Mignani07} for a discussion of this discrepancy). However, our measurements are of greater precision than those previously reported owing to the longer integration time and better spatial resolution of the {\em HST} observations. Thus, we obtained the first fully independent, and most accurate measurement of the Vela pulsar polarisation in the optical.  We could see no evidence of significant variation of the pulsar polarisation over the 5 days covered by the {\em HST} observations (Table 1), suggesting that the optical polarisation of pulsars does not vary over short time scales. This is also the case of the Crab, whose polarisation, also measured with the ACS, does not change up to time scales of a few weeks \citep{Moran13}. Nothing can be said for PSR\, B0540--69, the third pulsar for which optical polarisation has been measured by the {\em HST} \citep{Mignani10}. Indeed, in that case the polarisation measurement was obtained with the WFPC2 and, because of the different instrument observing strategy in polarisation mode, the data set did not consist of a series of repeated measurements at each polariser angle. Monitoring the polarisation of these pulsars on a regular time frame would be important to spot possible secular changes. In addition, phase-resolved polarisation measurements, so far carried out for the Crab pulsar only \citep{Slowikowska09}, would be important to track any change in the pulsar polarisation properties following a glitch, or in coincidence with giant pulses.


As in the case of \citet{Mignani07}, we have found an apparent alignment between the phase-averaged polarisation direction of the Vela pulsar ($\rm P.A.=146.3\degr\pm2.4\degr$), the axis of symmetry of the X-ray arcs and jets observed by {\em Chandra}, which are at position angle of 
$310\degr\pm1.5\degr$  \citep{Helfand01}, and its proper motion vector ($301\degr\pm1.8\degr$) measured with the VLBI \citep{Dodson03b}. This alignment is shown in Fig.\ 4,  where the orientation of these three vectors is compared with the phase-average polarisation direction measured by \citet{Mignani07} with the VLT.  Interestingly enough,  \citet{Moran13} have found the same scenario for the Crab pulsar. This suggests that the \lq\lq kick\rq\rq\ given to neutron stars at birth, at least in the case of the Crab and Vela pulsars, is directed along the rotation axis \citep{Lai01}, assuming that this is aligned with the proper motion vector. The alternative view is that the apparent alignment is an effect of projection onto the sky plane, and that there is no physical jet along the axis of rotation \citep{RD01}. More concrete measurements of the optical polarisation of pulsars will yield the needed observational restraints on these hypotheses. For the other best candidate, PSR\, B0540--69, very little can be said because the X-ray structure of the PWN is not clearly resolved by {\em Chandra} owing to the LMC distance. Moreover, the pulsar has no measured proper motion to compare with \citep{Mignani10}. Interestingly enough, however, the optical structure of the PWN, which broadly traces that of the X-ray PWN detected by {\em Chandra}, shows a possible alignment between its major axis and the pulsar polarisation vector. If the PSR\, B0540--69 PWN has indeed an arc-like structure along its major axis, this would suggest a possible different scenario with respect to both Crab and Vela.

\begin{figure}
\centering
\includegraphics[width=80mm]{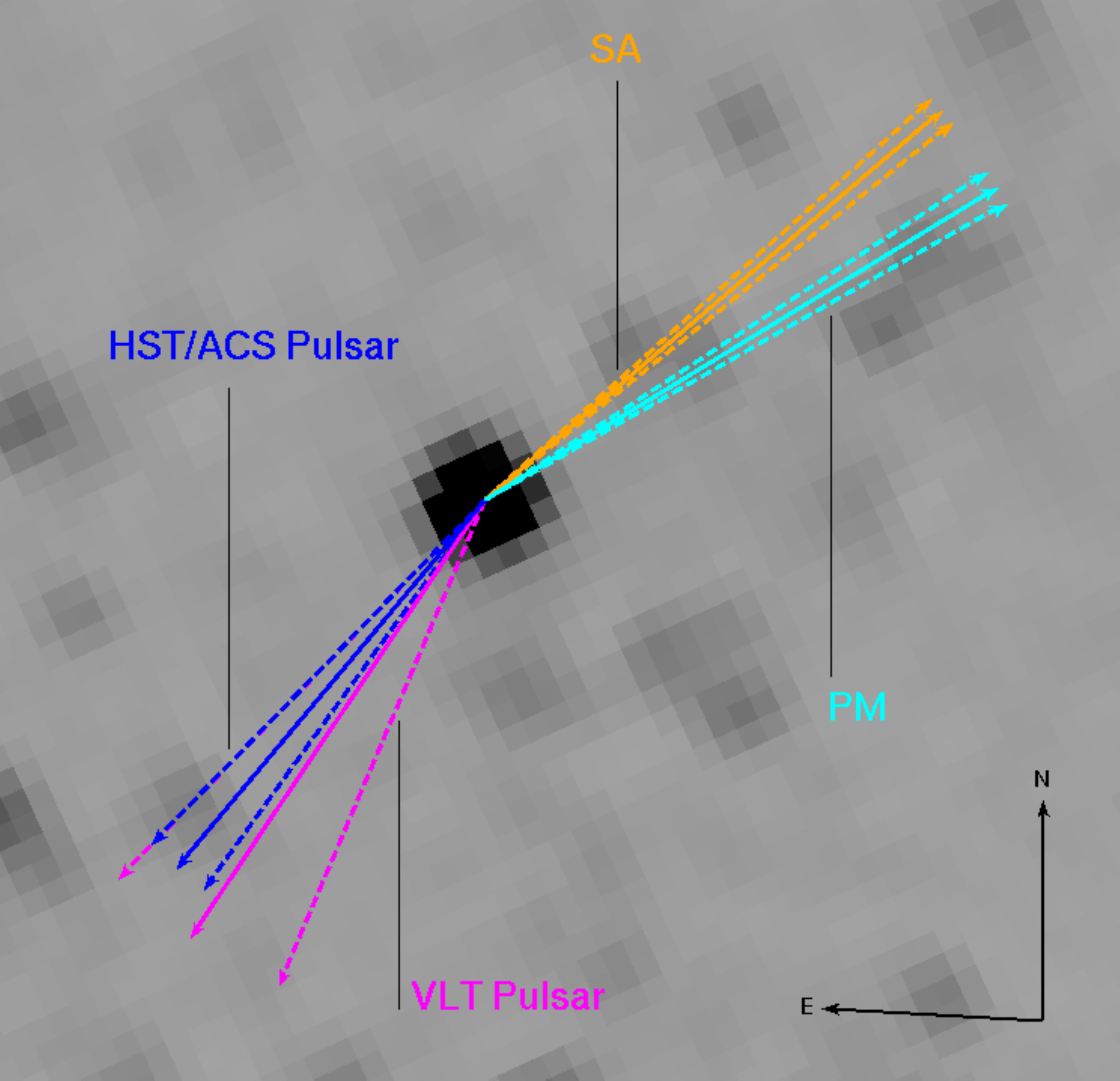}
\caption{ACS/WFC image of the Vela pulsar region (2011 February 18, FOV $\approx2\times2 \rm \ arcsec^{2}$). The vectors included are as follows: spin-axis vector (SA) ($310\degr\pm1.5\degr$; \citealt{Helfand01}), proper motion vector (PM) ($301.0\degr\pm1.8\degr$; \citealt{Dodson03b}), and the polarisation position angle of the pulsar ($146.3\degr\pm2.4\degr$) from this work. Also, included is the VLT measurement of the polarisation position angle of the pulsar ($146\degr\pm11\degr$; \citealt{Mignani07}). The dashed vectors denote the $1 \sigma$ uncertainties in the position angles.}
\label{figure4}
\end{figure}


Our improved measurement of the Vela pulsar polarisation does not affect the results presented in \citet{Mignani07} about the comparison with the expectations of various pulsar magnetosphere models.  For instance, using the same code as used in \citet{Mignani07} and  the same model parameters, the outer gap model \citep{Romani95} still predicts values of the phase-averaged P.D. that are typically much  larger than the  observed one, unless some depolarisation effects are introduced. In particular, our value of P.D., together with its currently associated error, makes it still difficult to set very accurate constraints on the dipole inclination angle $\alpha$, assuming a realistic viewing angle $\zeta \sim65\degr$ compatible with the profile of the pulsar optical light curve (e.g., \citealt{Gouiffes98}). This is because the values of the phase-averaged P.D. predicted by the outer gap model are much less dependent on $\alpha$ in the low polarisation regime (see, Fig.\ 3 of \citealt{Mignani07}).  According to our measured $\rm P.D.=8.1\% \pm0.6\%$ we can only say that the dipole inclination angle $\alpha$ would be probably close to $\approx 80\degr$. Similarly, the comparison of our measured $\rm P.A.=146.3\degr\pm2.4\degr$ with that predicted by the outer gap model would imply, for the same model parameters as before, that the position angle $\psi_0$ of the pulsarÕs projected rotational axis  would be $\approx130\degr$ (see, Fig.\ 4 of \citealt{Mignani07}). Modulo $180\degr$, this value is strikingly close to the position angles of the axis of symmetry of the X-ray arcs and jets ($310\degr\pm1.5\degr$; \citealt{Helfand01}) and of the proper motion vector ($301\degr\pm1.8\degr$;  \citealt{Dodson03b}).  



Examining the polarisation vector maps (see Fig. \ref{figure1}), one can see that both the levels of linear polarisation and position angles in the vicinity of the pulsar, including the inner part of the X-ray PWN region, are not much different from those of the rest of the field.  In other words, the sky around the pulsar is  more or less uniformly polarised. In order to quantify the mean sky polarisation properties in the Vela pulsar field we built the  histograms of the distributions of

P.D. and P.A., which we show in Figs.\ 5 and 6, respectively. The values represented in the histograms are those extracted from the polarisation map (Fig. \ref{figure1}), and are computed  over cells of $\approx 0.8 \times 0.8$ arcsec$^2$, evenly distributed in the FOV and far from the regions at the edge of the detector, which are affected by vignetting.
From the histogram of the P.D. distribution we see that, on average,  the sky is more strongly polarised than the pulsar, with a peak value of $\rm P.D. \approx 12\%$.  Furthermore, from the histogram of the P.A. distribution we also see that the polarisation P.A. of the pulsar ($\sim 140\degr$) is away from the peak of the sky distribution

($\sim 71\degr$). This shows that the polarisation properties of the pulsar are different from those of the rest of the field. 

From analysis of the geometry of the HST/ACS pointing (via the header files), we found that the plane of the scattering is perpendicular to the mean of the distribution of polarisation position angles in the pulsar field ($\sim$ 71$^{\circ}$) (see Figs. 6 and 7). Hence, this, together with the sky brightness ($\approx$ 23.6 magnitudes arcsec$^{-2}$), indicates that these observations are affected by zodiacal light. Furthermore, it suggests that the high level of sky polarisation is mostly due to zodiacal light rather than the nebula.

\begin{figure}
\centering
\includegraphics[width=85mm]{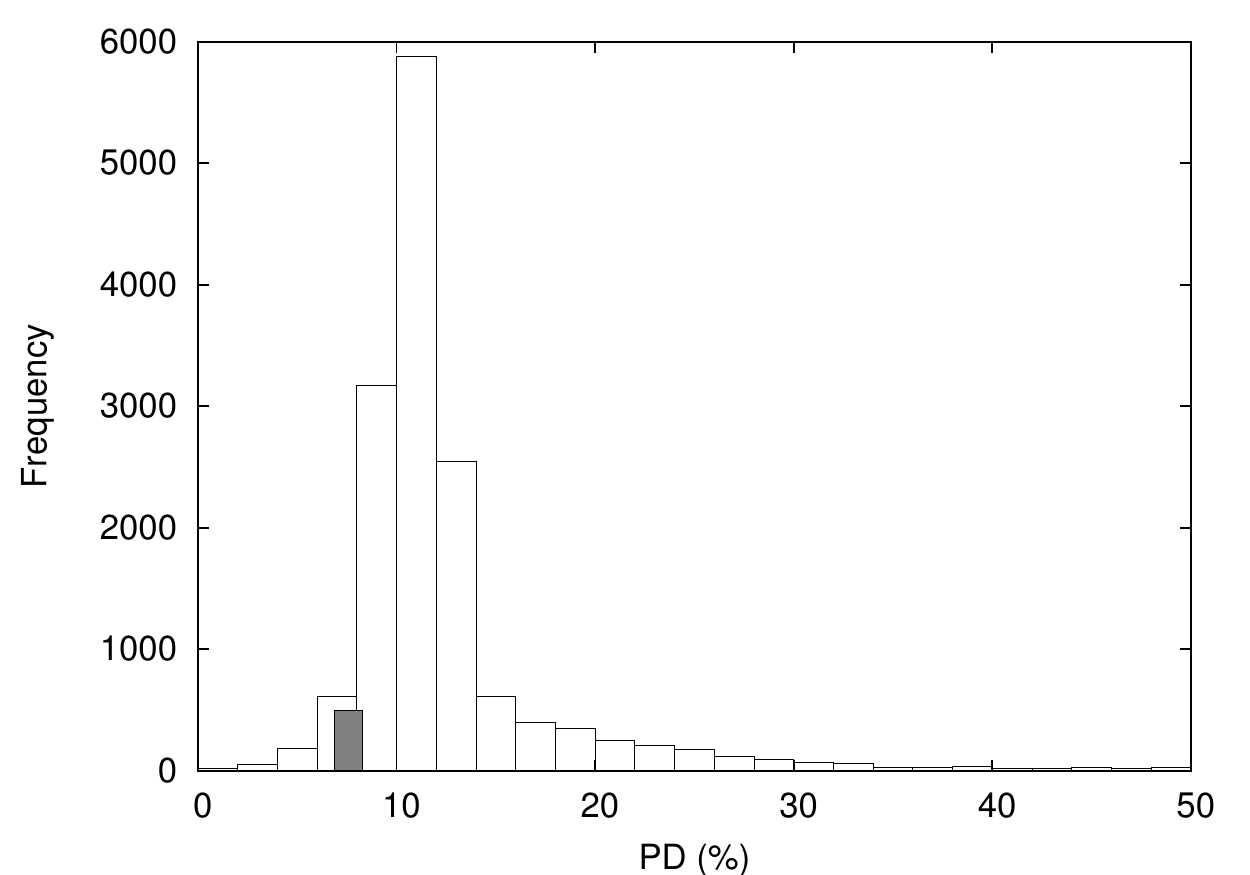}
\caption{Histogram of the observed P.D. (\%) measured in the sky background around the Vela pulsar. The size of the cells used for the mapping are $\approx 0.8 \times 0.8$ arcsec$^2$.
 The filled column denotes the P.D. of the pulsar.}
\label{figure5}
\end{figure}

\begin{figure}
\centering
\includegraphics[width=85mm]{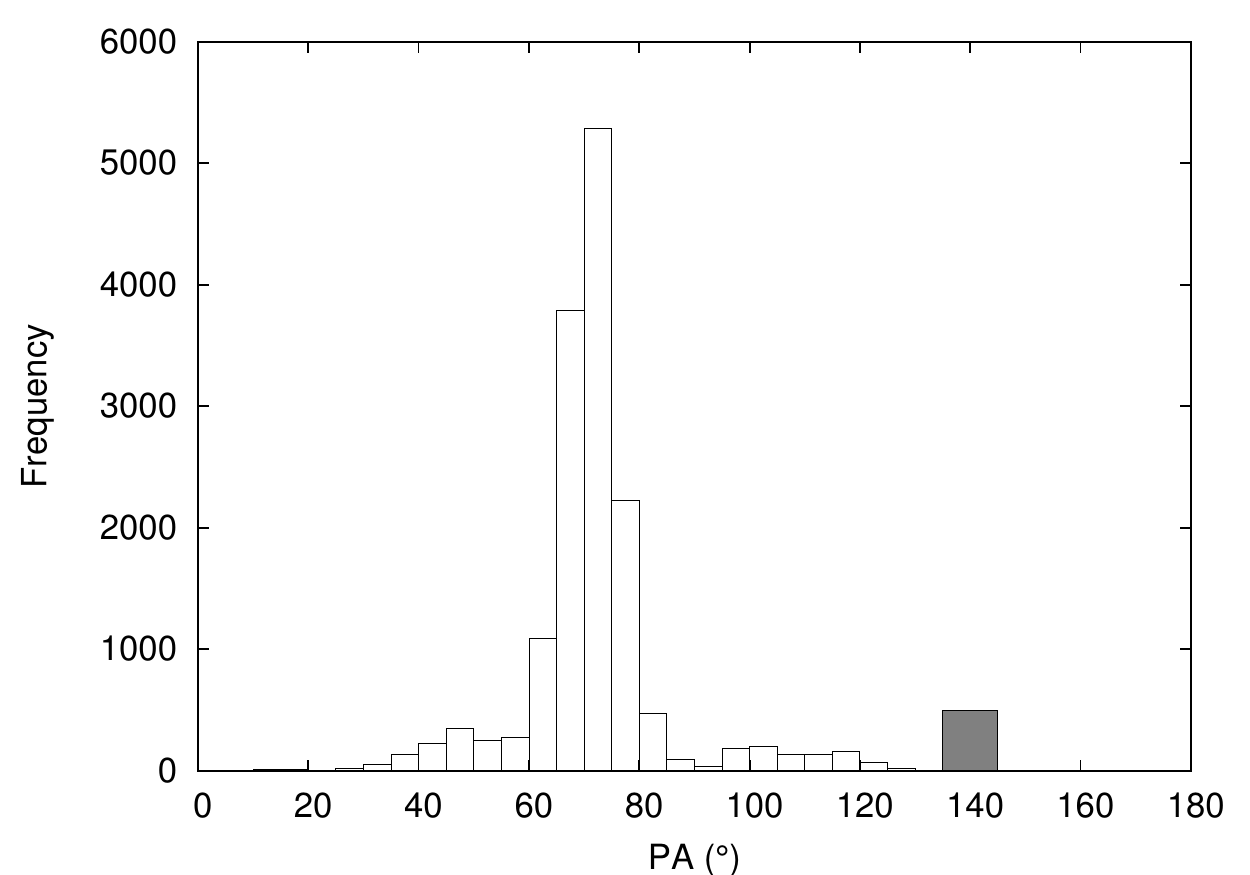}
\caption{Histogram of the observed P.A. ($\degr$) measured in the  sky background around the Vela pulsar. The size of the cells used for the mapping are $\approx 0.8 \times 0.8$ arcsec$^2$.
The filled column denotes the P.A. of the pulsar.}
\label{figure6}
\end{figure}

\begin{figure}
\centering
\includegraphics[width=70mm]{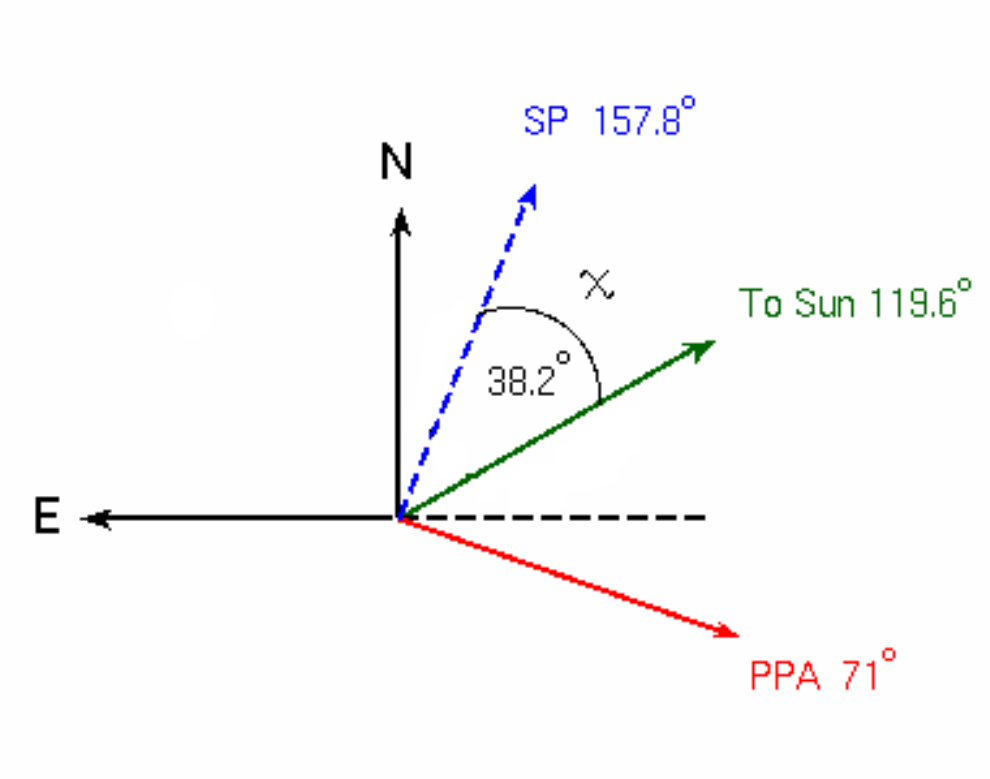}
\caption{The polarisation position angle (PPA) ($\sim$ 71$^{\circ}$) is perpendicular to the plane of scattering (SP) (157.8$^{\circ}$). This suggests that the sky background polarisation is due to zodiacal light rather than the nebula.}
\label{figure6}
\end{figure}

The non-detection of the Vela PWN even in deep {\em HST} exposures means that it is intrinsically much fainter in the optical than the Crab and PSR B0540-69 PWNe. 
This can be due to the fact that either the optical brightness of PWNe is not uniquely related to the pulsar spin-down power, with some pulsars injecting larger fractions of their spin-down power in the acceleration of relativistic particles that powers the PWN emission, and/or it is not uniquely related to the PWN X-ray luminosity.

Finally, it is also possible that the low optical surface brightness of the Vela PWN is also affected by the different physical conditions in the ejecta of the surrounding SNR, such as the local density, and/or by the intensity and properties of its magnetic field, which might lead to a different confinement of the pulsar relativistic wind. A better characterisation of the SNR environment in the proximity of the pulsar would be crucial to verify this possibility.


\section{Conclusions}

We have studied the phase-averaged polarisation properties of the Vela pulsar  using archival {\em HST}/ACS data covering a time span of 5 days. Our work marks the first high-spatial resolution multi-epoch study of the polarisation properties of the pulsar.   We found that the pulsar is polarised, with  $\rm P.D.=8.1\%\pm0.6\%$ and $\rm P.A.=146.3\degr\pm2.4\degr$, and that its polarisation properties do not change significantly over the short time span covered by the {\em HST} observations.  Our measurement independently confirms those obtained by \citet{Wagner00} and \citet{Mignani07}, using the same VLT data set, but are of greater precision. Thus, we confirm that important depolarisation factors need to be taken into account to make the measured polarisation value consistent with the expectation of most pulsar magnetosphere models, such at the outer gap model.  Future phase-resolved optical polarisation observations of the Vela pulsar, never performed so far, will bring more information on the geometry of the pulsar emission regions, crucial for a better comparison with theoretical models, together with the development of more advanced simulation codes. For example, \citet {McDonald11} have developed an inverse mapping approach for determining the emission height of the optical photons from pulsars. It uses the optical Stokes parameters to determine the most likely geometry for emission, including: magnetic field inclination angle ($\alpha$), the observers line of sight angle ($\zeta$), and emission height.

As in the case of \citet{Mignani07}, we found an apparent alignment between the position angle of the Vela pulsar phase-averaged polarisation vector, the axis of symmetry of the arcs of the X-ray PWN,  and the proper motion vector and spin-axis vector, as observed in  the Crab PWN \citep{Moran13}. Whether this characteristic is unique to all young pulsar/PWNe systems cannot be determined at the moment, owing to the lack of a sufficiently representative sample, with Vela being only the third young pulsar with a detected X-ray PWN, a measured pulsar proper motion, and a determined  phase-average polarisation direction.  

Finally, we present the first and deepest polarised images of the environment surrounding the Vela pulsar.  We found that the PWN is undetected in polarised light as is the case in unpolarised light, down to a limit of 26.8 magnitudes arcsec$^{-2}$, not quite as deep as  that obtained by \citet{Mignani03} (28.1 magnitudes arcsec$^{-2}$) with the WFPC2. 

The intrinsic faintness of the Vela PWN with respect to the Crab cannot be easily explained in terms of the pulsar energetics and lower surface brightness of the X-ray PWN, assuming that the X-ray and optical PWN brightnesses are uniquely related to each other, but is probably associated with a dramatic spectral turnover of the PWN between the X rays and the optical.

From a general stand point, more multi-wavelength polarisation observations of pulsars, both phase-averaged and phase-resolved, and of their PWNe with existing instruments, such as the  {\em HST}/ACS and GASP in the optical or {\em INTEGRAL}/IBIS in soft $\gamma$ rays, and instruments aboard future  X-ray missions such as the {\em Gravity and Extreme Magnetism  Small Explorer} ({\em GEMS}; \citealt{Ghosh13}) or the {\em X-ray Imaging Polarimetry Explorer} ({\em XIPE}; \citealt{Soffitta13}) in the X-rays

will help to provide the much needed data to constrain current theoretical models of emission from pulsar magnetospheres. 


\section*{Acknowledgments} 

All of the data presented in this paper were obtained from the Mikulski Archive for Space Telescopes (MAST). STScI is operated by the Association of Universities for Research in Astronomy, Inc., under NASA contract NAS5-26555. Support for MAST for non-HST data is provided by the NASA Office of Space Science via grant NNX09AF08G and by other grants and contracts. We acknowledge the use of data, for the X-ray contours of the Vela PWN, obtained from the Chandra Data Archive and the Chandra Source Catalog. We thank Jeremy Walsh, ESO, for the use of his polarimetry software IMPOL to produce the polarisation maps. PM is grateful for his funding from the Irish Research Council (IRC). RPM thanks the European Commission Seventh Framework Programme (FP7/2007-2013) for their support under grant agreement n.267251. This paper is partially financed by the grant DEC-2012/05/B/ST9/03924 of the Polish National Science Centre. The anonymous referee is thanked for comments which significantly improved the original version of the paper


\bibliographystyle{mn2e}
\bibliography{reference_database}

\begin{thebibliography}{}

\bibitem[\protect\citeauthoryear{{Caraveo}, {De Luca}, {Mignani} \&
  {Bignami}}{{Caraveo} et~al.}{2001}]{Caraveo01}
{Caraveo} P.~A.,  {De Luca} A.,  {Mignani} R.~P.,    {Bignami} G.~F.,  2001,
  ApJ, 561, 930

\bibitem[\protect\citeauthoryear{{Chanan} \& {Helfand}}{{Chanan} \&
  {Helfand}}{1990}]{Chanan90}
{Chanan} G.~A.,  {Helfand} D.~J.,  1990, ApJ, 352, 167

\bibitem[\protect\citeauthoryear{{Collins}, {Kyne}, {Lara}, {Redfern},
  {Shearer} \& {Sheehan}}{{Collins} et~al.}{2013}]{Collins13}
{Collins} P.,  {Kyne} G.,  {Lara} D.,  {Redfern} M.,  {Shearer} A.,
  {Sheehan} B.,  2013, Experimental Astronomy, 36, 479

\bibitem[\protect\citeauthoryear{{Dean}, {Clark}, {Stephen}, {McBride},
  {Bassani}, {Bazzano}, {Bird}, {Hill}, {Shaw} \& {Ubertini}}{{Dean}
  et~al.}{2008}]{Dean08}
{Dean} A.~J.,  {Clark} D.~J.,  {Stephen} J.~B.,  {McBride} V.~A.,  {Bassani}
  L.,  {Bazzano} A.,  {Bird} A.~J.,  {Hill} A.~B.,  {Shaw} S.~E.,    {Ubertini}
  P.,  2008, Science, 321, 1183

\bibitem[\protect\citeauthoryear{{Dodson}, {Legge}, {Reynolds} \&
  {McCulloch}}{{Dodson} et~al.}{2003}]{Dodson03b}
{Dodson} R.,  {Legge} D.,  {Reynolds} J.~E.,    {McCulloch} P.~M.,  2003, ApJ,
  596, 1137

\bibitem[\protect\citeauthoryear{{Dodson}, {Lewis}, {McConnell} \&
  {Deshpande}}{{Dodson} et~al.}{2003}]{Dodson03a}
{Dodson} R.,  {Lewis} D.,  {McConnell} D.,    {Deshpande} A.~A.,  2003, MNRAS,
  343, 116

\bibitem[\protect\citeauthoryear{{Dombrovsky}}{{Dombrovsky}}{1954}]{Dombrovsky54}
{Dombrovsky} V.~A.,  1954, Doklady Akad Nauk, USSR, 94, 1021

\bibitem[\protect\citeauthoryear{{Durant}, {Kargaltsev}, {Pavlov}, {Kropotina}
  \& {Levenfish}}{{Durant} et~al.}{2013}]{Durant13}
{Durant} M.,  {Kargaltsev} O.,  {Pavlov} G.~G.,  {Kropotina} J.,    {Levenfish}
  K.,  2013, ApJ, 763, 72

\bibitem[\protect\citeauthoryear{{Forot}, {Laurent}, {Grenier}, {Gouiff{\`e}s}
  \& {Lebrun}}{{Forot} et~al.}{2008}]{Forot08}
{Forot} M.,  {Laurent} P.,  {Grenier} I.~A.,  {Gouiff{\`e}s} C.,    {Lebrun}
  F.,  2008, ApJL, 688, L29

\bibitem[\protect\citeauthoryear{{Gaensler} \& {Slane}}{{Gaensler} \&
  {Slane}}{2006}]{Gaensler06}
{Gaensler} B.~M.,  {Slane} P.~O.,  2006, ARA\&A, 44, 17

\bibitem[\protect\citeauthoryear{{Ghosh}, {Angelini}, {Baring}, {Baumgartner},
  {Black} \& {et al.}}{{Ghosh} et~al.}{2013}]{Ghosh13}
{Ghosh} P.,  {Angelini} L.,  {Baring} M.,  {Baumgartner} W.,  {Black} K.,
  {et al.} 2013, ArXiv e-prints

\bibitem[\protect\citeauthoryear{{Gouiffes}}{{Gouiffes}}{1998}]{Gouiffes98}
{Gouiffes} C.,  1998, in {Shibazaki} N.,  ed., Neutron Stars and Pulsars:
  Thirty Years after the Discovery {Optical Observation of the VELA Pulsar}.
p.~363

\bibitem[\protect\citeauthoryear{{Helfand}, {Gotthelf} \& {Halpern}}{{Helfand}
  et~al.}{2001}]{Helfand01}
{Helfand} D.~J.,  {Gotthelf} E.~V.,    {Halpern} J.~P.,  2001, ApJ, 556, 380

\bibitem[\protect\citeauthoryear{{Kargaltsev} \& {Pavlov}}{{Kargaltsev} \&
  {Pavlov}}{2008}]{Kargaltsev08}
{Kargaltsev} O.,  {Pavlov} G.~G.,  2008, in {Bassa} C.,  {Wang} Z.,  {Cumming}
  A.,   {Kaspi} V.~M.,  eds, 40 Years of Pulsars: Millisecond Pulsars,
  Magnetars and More Vol.~983 of American Institute of Physics Conference
  Series, {Pulsar Wind Nebulae in the Chandra Era}.
pp 171--185

\bibitem[\protect\citeauthoryear{{Kargaltsev}, {Rangelov} \&
  {Pavlov}}{{Kargaltsev} et~al.}{2013}]{Kargaltsev13}
{Kargaltsev} O.,  {Rangelov} B.,    {Pavlov} G.~G.,  2013, ArXiv e-prints

\bibitem[\protect\citeauthoryear{{Kern}, {Martin}, {Mazin} \& {Halpern}}{{Kern}
  et~al.}{2003}]{Kern03}
{Kern} B.,  {Martin} C.,  {Mazin} B.,    {Halpern} J.~P.,  2003, ApJ, 597, 1049

\bibitem[\protect\citeauthoryear{{Kristian}, {Visvanathan}, {Westphal} \&
  {Snellen}}{{Kristian} et~al.}{1970}]{Kristian70}
{Kristian} J.,  {Visvanathan} N.,  {Westphal} J.~A.,    {Snellen} G.~H.,  1970,
  ApJ, 162, 475

\bibitem[\protect\citeauthoryear{{Kyne}, {Sheehan}, {Collins}, {Redfern} \&
  {Shearer}}{{Kyne} et~al.}{2010}]{Kyne10}
{Kyne} G.,  {Sheehan} B.,  {Collins} P.,  {Redfern} M.,    {Shearer} A.,  2010,
  in European Physical Journal Web of Conferences Vol.~5 of European Physical
  Journal Web of Conferences, {GASP-Galway astronomical Stokes polarimeter}.
p.~5003

\bibitem[\protect\citeauthoryear{{Lai}, {Chernoff} \& {Cordes}}{{Lai}
  et~al.}{2001}]{Lai01}
{Lai} D.,  {Chernoff} D.~F.,    {Cordes} J.~M.,  2001, ApJ, 549, 1111

\bibitem[\protect\citeauthoryear{{Large}, {Vaughan} \& {Mills}}{{Large}
  et~al.}{1968}]{Large68}
{Large} M.~I.,  {Vaughan} A.~E.,    {Mills} B.~Y.,  1968, Nature, 220, 340

\bibitem[\protect\citeauthoryear{{Markwardt} \& {{\"O}gelman}}{{Markwardt} \&
  {{\"O}gelman}}{1995}]{Markwardt95}
{Markwardt} C.~B.,  {{\"O}gelman} H.,  1995, Nature, 375, 40

\bibitem[\protect\citeauthoryear{{McDonald}, {O'Connor}, {de Burca}, {Golden}
  \& {Shearer}}{{McDonald} et~al.}{2011}]{McDonald11}
{McDonald} J.,  {O'Connor} P.,  {de Burca} D.,  {Golden} A.,    {Shearer} A.,
  2011, MNRAS, 417, 730

\bibitem[\protect\citeauthoryear{{Middleditch}, {Pennypacker} \&
  {Burns}}{{Middleditch} et~al.}{1987}]{Middleditch87}
{Middleditch} J.,  {Pennypacker} C.~R.,    {Burns} M.~S.,  1987, ApJ, 315, 142

\bibitem[\protect\citeauthoryear{{Mignani}}{{Mignani}}{2011}]{Mignani11}
{Mignani} R.~P.,  2011, Advances in Space Research, 47, 1281

\bibitem[\protect\citeauthoryear{{Mignani}, {Bagnulo}, {Dyks}, {Lo Curto} \&
  {S{\l}owikowska}}{{Mignani} et~al.}{2007}]{Mignani07}
{Mignani} R.~P.,  {Bagnulo} S.,  {Dyks} J.,  {Lo Curto} G.,    {S{\l}owikowska}
  A.,  2007, A\&AP, 467, 1157

\bibitem[\protect\citeauthoryear{{Mignani} \& {Caraveo}}{{Mignani} \&
  {Caraveo}}{2001}]{Mignani01}
{Mignani} R.~P.,  {Caraveo} P.~A.,  2001, A\&AP, 376, 213

\bibitem[\protect\citeauthoryear{{Mignani}, {De Luca}, {Kargaltsev}, {Pavlov},
  {Zaggia}, {Caraveo} \& {Becker}}{{Mignani} et~al.}{2003}]{Mignani03}
{Mignani} R.~P.,  {De Luca} A.,  {Kargaltsev} O.,  {Pavlov} G.~G.,  {Zaggia}
  S.,  {Caraveo} P.~A.,    {Becker} W.,  2003, ApJ, 594, 419

\bibitem[\protect\citeauthoryear{{Mignani}, {Sartori}, {de Luca}, {Rudak},
  {S{\l}owikowska}, {Kanbach} \& {Caraveo}}{{Mignani} et~al.}{2010}]{Mignani10}
{Mignani} R.~P.,  {Sartori} A.,  {de Luca} A.,  {Rudak} B.,  {S{\l}owikowska}
  A.,  {Kanbach} G.,    {Caraveo} P.~A.,  2010, A\&AP, 515, A110

\bibitem[\protect\citeauthoryear{{Moran}, {Shearer}, {Mignani},
  {S{\l}owikowska}, {De Luca}, {Gouiff{\`e}s} \& {Laurent}}{{Moran}
  et~al.}{2013}]{Moran13}
{Moran} P.,  {Shearer} A.,  {Mignani} R.~P.,  {S{\l}owikowska} A.,  {De Luca}
  A.,  {Gouiff{\`e}s} C.,    {Laurent} P.,  2013, MNRAS, 433, 2564

\bibitem[\protect\citeauthoryear{{Nasuti}, {Mignani}, {Caraveo} \&
  {Bignami}}{{Nasuti} et~al.}{1996}]{Nasuti96}
{Nasuti} F.~P.,  {Mignani} R.,  {Caraveo} P.~A.,    {Bignami} G.~F.,  1996,
  aap, 314, 849

\bibitem[\protect\citeauthoryear{{Pavlov}, {Teter}, {Kargaltsev} \&
  {Sanwal}}{{Pavlov} et~al.}{2003}]{Pavlov03}
{Pavlov} G.~G.,  {Teter} M.~A.,  {Kargaltsev} O.,    {Sanwal} D.,  2003, ApJ,
  591, 1157

\bibitem[\protect\citeauthoryear{{Pavlov}, {Zavlin}, {Sanwal}, {Burwitz} \&
  {Garmire}}{{Pavlov} et~al.}{2001}]{Pavlov01b}
{Pavlov} G.~G.,  {Zavlin} V.~E.,  {Sanwal} D.,  {Burwitz} V.,    {Garmire}
  G.~P.,  2001, ApJL, 552, L129

\bibitem[\protect\citeauthoryear{{Pavlovsky}}{{Pavlovsky}}{2004}]{Pavlovsky04}
{Pavlovsky} C.,  2004, ACS Data Handbook, 3, 3.0

\bibitem[\protect\citeauthoryear{{Radhakrishnan} \&
  {Deshpande}}{{Radhakrishnan} \& {Deshpande}}{2001}]{RD01}
{Radhakrishnan} V.,  {Deshpande} A.~A.,  2001, A\&AP, 379, 551

\bibitem[\protect\citeauthoryear{{Romani} \& {Yadigaroglu}}{{Romani} \&
  {Yadigaroglu}}{1995}]{Romani95}
{Romani} R.~W.,  {Yadigaroglu} I.-A.,  1995, ApJ, 438, 314

\bibitem[\protect\citeauthoryear{{Shearer}}{{Shearer}}{2008}]{Shearer08}
{Shearer} A.,  2008, in {Phelan} D.,  {Ryan} O.,   {Shearer} A.,  eds,
  Astrophysics and Space Science Library Vol.~351 of Astrophysics and Space
  Science Library, {High Time Resolution Astrophysics and Pulsars}.
pp 1--4020

\bibitem[\protect\citeauthoryear{{Shklovsky}}{{Shklovsky}}{1953}]{Shklovsky53}
{Shklovsky} I.~S.,  1953, Doklady Akad Nauk, USSR, 8, 135

\bibitem[\protect\citeauthoryear{{Simmons} \& {Stewart}}{{Simmons} \&
  {Stewart}}{1985}]{Simmons85}
{Simmons} J.~F.~L.,  {Stewart} B.~G.,  1985, A\&A, 142, 100

\bibitem[\protect\citeauthoryear{{Skrutskie}, {Cutri}, {Stiening}, {Weinberg},
  {Schneider} \& {et al.,}}{{Skrutskie} et~al.}{2006}]{Skrutskie06}
{Skrutskie} M.~F.,  {Cutri} R.~M.,  {Stiening} R.,  {Weinberg} M.~D.,
  {Schneider} S.,    {et al.,} 2006, ApJ, 131, 1163

\bibitem[\protect\citeauthoryear{{S{\l}owikowska}, {Kanbach}, {Kramer} \&
  {Stefanescu}}{{S{\l}owikowska} et~al.}{2009}]{Slowikowska09}
{S{\l}owikowska} A.,  {Kanbach} G.,  {Kramer} M.,    {Stefanescu} A.,  2009,
  MNRAS, 397, 103

\bibitem[\protect\citeauthoryear{{Smith}, {Jones}, {Dick} \& {Pike}}{{Smith}
  et~al.}{1988}]{Smith88}
{Smith} F.~G.,  {Jones} D.~H.~P.,  {Dick} J.~S.~B.,    {Pike} C.~D.,  1988,
  MNRAS, 233, 305

\bibitem[\protect\citeauthoryear{{Soffitta}, {Barcons}, {Bellazzini}, {Braga},
  {Costa} \& {et al.}}{{Soffitta} et~al.}{2013}]{Soffitta13}
{Soffitta} P.,  {Barcons} X.,  {Bellazzini} R.,  {Braga} J.,  {Costa} E.,
  {et al.} 2013, Experimental Astronomy, 36, 523

\bibitem[\protect\citeauthoryear{{Sparks} \& {Axon}}{{Sparks} \&
  {Axon}}{1999}]{Sparks99}
{Sparks} W.~B.,  {Axon} D.~J.,  1999, PASP, 111, 1298

\bibitem[\protect\citeauthoryear{{Vashakidze}}{{Vashakidze}}{1954}]{Vashakidze54}
{Vashakidze} M.~A.,  1954, Astron. Circ, 147, 11

\bibitem[\protect\citeauthoryear{{Wagner} \& {Seifert}}{{Wagner} \&
  {Seifert}}{2000}]{Wagner00}
{Wagner} S.~J.,  {Seifert} W.,  2000, in {Kramer} M.,  {Wex} N.,
  {Wielebinski} R.,  eds, IAU Colloq. 177: Pulsar Astronomy - 2000 and Beyond
  Vol.~202 of Astronomical Society of the Pacific Conference Series, {Optical
  Polarization Measurements of Pulsars}.
p.~315

\bibitem[\protect\citeauthoryear{{Walsh}}{{Walsh}}{1999}]{Walsh99}
{Walsh} J.,  1999, Issue Number 26

\bibitem[\protect\citeauthoryear{{Wampler}, {Scargle} \& {Miller}}{{Wampler}
  et~al.}{1969}]{Wampler69}
{Wampler} E.~J.,  {Scargle} J.~D.,    {Miller} J.~S.,  1969, ApJL, 157, L1

\bibitem[\protect\citeauthoryear{{Weisskopf}, {Hester}, {Tennant}, {Elsner},
  {Schulz}, {Marshall}, {Karovska}, {Nichols}, {Swartz}, {Kolodziejczak} \&
  {O'Dell}}{{Weisskopf} et~al.}{2000}]{Weisskopf00}
{Weisskopf} M.~C.,  {Hester} J.~J.,  {Tennant} A.~F.,  {Elsner} R.~F.,
  {Schulz} N.~S.,  {Marshall} H.~L.,  {Karovska} M.,  {Nichols} J.~S.,
  {Swartz} D.~A.,  {Kolodziejczak} J.~J.,    {O'Dell} S.~L.,  2000, ApJL, 536,
  L81

\bibitem[\protect\citeauthoryear{{Weisskopf}, {Silver}, {Kestenbaum}, {Long} \&
  {Novick}}{{Weisskopf} et~al.}{1978}]{Weisskopf78}
{Weisskopf} M.~C.,  {Silver} E.~H.,  {Kestenbaum} H.~L.,  {Long} K.~S.,
  {Novick} R.,  1978, ApJL, 220, L117

\end{thebibliography}


\label{lastpage}

\end{document}